\documentclass[a4paper,11pt,preprintnumbers]{article}
\pdfoutput=1

\usepackage{physics}
\usepackage{makecell}
\usepackage{amssymb,mathrsfs,amsmath}
\usepackage{a4wide}
\usepackage{color}
\usepackage[dvipsnames,table]{xcolor}
\usepackage{slashed,soul}
\usepackage{graphicx}
\usepackage{amsfonts}
\usepackage{lscape}
\def\linkcolor{cyan!70!black}

\usepackage[
colorlinks=true
,urlcolor=\linkcolor
,anchorcolor=\linkcolor
,citecolor=\linkcolor
,filecolor=\linkcolor
,linkcolor=\linkcolor
,menucolor=\linkcolor
,linktocpage=true
,pdfproducer=medialab
,pdfa=true
]{hyperref}
\usepackage{amsthm}
\usepackage{booktabs}
\usepackage{array}
\usepackage{rotating}
\usepackage[numbers, sort&compress]{natbib}
\usepackage{multirow}
\usepackage{float}
\usepackage[utf8]{inputenc}
\usepackage[T1]{fontenc}
\usepackage{appendix}
\usepackage{epsfig}
\usepackage{orcidlink}
\usepackage{comment}
\usepackage{adjustbox}
\usepackage{tcolorbox}
\usepackage{booktabs}

\newcommand{\beq}{\begin{equation}} 
\newcommand{\eeq}{\end{equation}} 
\newcommand{\ba}{\begin{array}}  
\newcommand{\ea}{\end{array}} 
\newcommand{\bea}{\begin{eqnarray}}  
\newcommand{\eea}{\end{eqnarray} }  
\newcommand{\bal}{\begin{align}}
\newcommand{\eal}{\end{align}}   
\newcommand{\bi}{\begin{itemize}}  
\newcommand{\ei}{\end{itemize}}  
\newcommand{\ben}{\begin{enumerate}}  
\newcommand{\een}{\end{enumerate}}  
\newcommand{\bc}{\begin{center}}
\newcommand{\ec}{\end{center}} 
\newcommand{\bt}{\begin{table}}
\newcommand{\et}{\end{table}}  
\newcommand{\btb}{\begin{tabular}}
\newcommand{\etb}{\end{tabular}}

\allowdisplaybreaks

\let\OLDthebibliography\thebibliography
\renewcommand\thebibliography[1]{
  \OLDthebibliography{#1}
  \setlength{\parskip}{0pt}
  \setlength{\itemsep}{0pt plus 0.3ex}
}

\begin{document}

\vspace{1cm}

\begin{titlepage}

\begin{flushright}
IFT-UAM/CSIC-24-108\\
 \end{flushright}
\vspace{0.2truecm}

\begin{center}
\renewcommand{\baselinestretch}{1.8}\normalsize
\boldmath
{\LARGE\textbf{
Completing the one-loop \\ $\nu$SMEFT Renormalization Group Evolution 
}}
\unboldmath
\end{center}

\vspace{0.4truecm}

\renewcommand*{\thefootnote}{\fnsymbol{footnote}}

\begin{center}

{
Marco Ardu$^1$\footnote{\href{mailto:marco.ardu@ific.uv.es}{marco.ardu@ific.uv.es}}\orcidlink{0000-0003-2352-0845}
and Xabier Marcano$^2$\footnote{\href{mailto:xabier.marcano@uam.es}{xabier.marcano@uam.es}}\orcidlink{0000-0003-0033-0504}
}

\vspace{0.7truecm}

{\footnotesize
$^1$ Instituto de F\'{\i}sica Corpuscular, Universidad de Valencia and CSIC\\ 
 Edificio Institutos Investigaci\'on, Catedr\'atico Jos\'e Beltr\'an 2, 46980 Spain
 \\[.5ex]
$^2$
Departamento de F\'{\i}sica Te\'orica and Instituto de F\'{\i}sica Te\'orica UAM/CSIC,\\
Universidad Aut\'onoma de Madrid, Cantoblanco, 28049 Madrid, Spain
}

\vspace*{2mm}
\end{center}

\renewcommand*{\thefootnote}{\arabic{footnote}}
\setcounter{footnote}{0}

\begin{abstract}

In this work we consider the Standard Model Effective Field Theory extended with right-handed neutrinos, the $\nu$SMEFT, and calculate the full set of one-loop anomalous dimensions that are proportional to Yukawa couplings.
These contributions are particularly relevant when symmetry-protected low scale seesaw models are embeded in the SMEFT, since large neutrino Yukawa couplings are expected.  
By combining our results with the already available gauge anomalous dimensions, we provide the complete set of one-loop renormalization group evolution equations for the {dimension six} $\nu$SMEFT. 
As a possible phenomenological implication of our results, we discuss the sensitivity of lepton flavor-violating observables to $\nu$SMEFT operators, focusing on the more sensitive $\mu\to e$ transitions.
 
\end{abstract}

\end{titlepage}

\tableofcontents

\section{Introduction}

Despite the great success of the Standard Model (SM) of Particle Physics, there are still open questions that need to be addressed, introducing the need to go Beyond the SM (BSM).
This has motivated a plethora of new theoretical models, usually predicting the existence of new particles that have been searched for at numerous experiments. 
Unfortunately, no experiment has been able to discover any of these new particles, leaving us with two  possibilities: either the new particles are very heavy, or they interact very feebly with the SM. 
The former invites us to consider an effective field theory (EFT) framework, such as the SMEFT~\cite{Buchmuller:1985jz,Grzadkowski:2010es}, which is gaining popularity due to the lack of positive signals at the LHC and as it allows for model-agnostic analyses.  
The latter usually considers specific models with new light but weakly-interacting particles, such as models to explain the origin of dark matter or the generation of neutrino masses. 
Here we consider both hypothesis at the same time, adding to the usual EFT framework new light degrees of freedom associated to neutrino mass generation mechanisms. 

Among the many UV-complete models for generating neutrino masses~\cite{Cai:2017jrq}, many of them introduce new massive almost-sterile neutrinos, usually known as sterile neutrinos, right-handed neutrinos or heavy neutral leptons (HNLs).
The simplest of these models is the type-I seesaw model~\cite{Minkowski:1977sc,Mohapatra:1979ia,Yanagida:1979as,Gell-Mann:1979vob}, although its canonical formulation becomes almost impossible to probe since it requires either too heavy neutrino masses or tiny Yukawa couplings. 
Fortunately, symmetry protected scenarios such as the inverse~\cite{Mohapatra:1986aw,Mohapatra:1986bd} or linear~\cite{Akhmedov:1995ip,Malinsky:2005bi} seesaw models allow for a low-energy realization of the model, where lighter neutrinos can still have large Yukawa couplings. 
This enriches the phenomenology of the models and, therefore, enhances its testability. 

Nevertheless, these low-scale seesaw models also open new theoretical questions.
On the one hand, we might wonder where this new symmetry protecting neutrino masses is coming from, including also the different patterns in the neutrino mass matrix distinguishing between the different low-scale realizations. 
On the other hand, having a model with new electroweak-scale masses and large Yukawa couplings worsens the Higgs stability problem~\cite{DiLuzio:2017tfn}.
These points call for an UV-completion of low scale seesaw models with not very heavy new dynamics, as has been studied for instance in Refs.~\cite{Ma:2009gu,Khalil:2010iu, Bazzocchi:2010dt,Okada:2012np, Law:2012mj,Basso:2012ti, Kajiyama:2012xg, Ma:2014qra, Fraser:2014yha, Wang:2015saa,Ahriche:2016acx,DeRomeri:2017oxa, Geng:2017foe, Nomura:2017kih, Mandal:2019oth,Mondal:2021vou,Fernandez-Martinez:2021ypo,Abada:2021yot,Arias-Aragon:2022ats,Batra:2023bqj,CarcamoHernandez:2023atk,Abada:2023zbb}.
In this work, we also assume that such completion exists, but we will instead parametrized it in an EFT framework, the so-called $\nu$SMEFT.

The $\nu$SMEFT considers the SMEFT extended with new massive neutrinos as light degrees of freedom. 
Complete non-redundant basis of operators up to dim-9 have been introduced in Refs.~\cite{delAguila:2008ir,Aparici:2009fh,Bhattacharya:2015vja,Liao:2016qyd, Li:2021tsq},
and the gauge contributions to the Renormalization Group Evolution equations (RGEs) of dim-6 operators have been calculated in \cite{Datta:2020ocb}{, see also~\cite{Chala:2020pbn}}. Moreover, the matching onto the low-energy version of the effective theory, commonly refer to as $\nu$LEFT, and which includes QED and QCD invariant operators with sterile neutrinos, is known at the tree level \cite{Matchingnusmeft}. 

The phenomenology of the $\nu$SMEFT operators received increasing attention over the last years~\cite{nusmeft1, nusmeft2,nusmeft3, nusmeft4,nusmeft5,nusmeft6,nusmeft7,nusmeft8,nusmeft9,nusmeft10,nusmeft11,nusmeft12,nusmeft13,nusmeft14, nusmeft15, nusmeft16, nusmeft17, nusmeft18, nusmeft19,nusmeft20,nusmeft21,nusmeft22,nusmeft23, nusmeft24, nusmeft25, nusmeft26, nusmeft27, nusmeft28, nusmeft29, nusmeft30,Fernandez-Martinez:2023phj, Fuyuto:2024oii}. However, these previous $\nu$SMEFT studies usually neglect the effect of the neutrino Yukawa couplings, as the standard assumption is to consider a single (Majorana) neutrino with a small mixing to the active ones, and study only the effects of the new EFT operators. 
{Among the few exceptions, Ref.~\cite{Datta:2021akg} did consider the neutrino Yukawa couplings and computed their contributions to the RGEs of 4-fermion operators.
Our goal here is to fill this gap, considering the effects of having sizable neutrino Yukawa couplings, as in the case of the low scale seesaw models. 
In particular, we study the complete set of contributions from Yukawa couplings to the $\nu$SMEFT RGEs, extending previous partial results~\cite{Chala:2020pbn,Datta:2020ocb,Datta:2021akg}, and arriving at the complete set of one-loop RGEs of this framework.
}

Moreover, as a particular phenomenological application of our computation, we focus on charged lepton flavor violating (LFV) transitions. 
Being forbidden in the SM, these processes provide one of our best opportunities to discover new physics (see~\cite{Calibbi:2017uvl, Ardu:2022sbt} for recent reviews), and as such they have been extensively studied in the EFT context~\cite{Raidal:1997hq,Kuno:1999jp,Brignole:2004ah,Cirigliano:2009bz,Crivellin:2013hpa,Celis:2014asa,Davidson:2017nrp, Cirigliano:2017azj, Crivellin:2017rmk, Husek:2020fru,Ardu:2021koz, Calibbi:2021pyh,Davidson:2022nnl,Calibbi:2022ddo,Hoferichter:2022mna,Fortuna:2022sxt,Ardu:2023yyw,Fortuna:2023paj, Plakias:2023esq, Ardu:2024bua, Fernandez-Martinez:2024bxg, Haxton:2024lyc}. 
Our new computation of Yukawa neutrino contribution to the RGEs allows for similar analysis also in the $\nu$SMEFT and, as we will show, provides new bounds to its operators. 

This paper is organized as follows. In Section \ref{sec:nuSMEFT} we define the notation and introduce our conventions for the operator basis. In Section \ref{sec:RGEs}, we briefly review the RGE derivation and describe the setup for our calculation. The complete results for the anomalous dimensions are presented in Appendix \ref{App:RGEs}. In Section \ref{sec:pheno} we illustrate an example of the phenomenological interest of our results, determining the sensitivity of $\mu\to e$ transitions  to $\nu$SMEFT operators by calculating their RGE mixing with charged lepton-flavour-changing operators. Finally, we draw our conclusions in Section \ref{sec:conclusions}.

\section[The $\nu$SMEFT]{The $\boldsymbol\nu$SMEFT}
\label{sec:nuSMEFT}
We augment the Standard Model with $n$ right-handed fermion gauge singlets, denoted as $N_a$, as well as all possible dimension 6 operators.
The resulting Lagrangian is given by
\begin{equation}
	\mathcal{L}=\mathcal{L}_{{\rm SM}+N_a}+\mathcal{L}^{(6)}\,,
\end{equation}
where
\begin{eqnarray}
	\mathcal{L}_{{\rm SM}+N_a}&=&-\frac{1}{4}B_{\mu\nu}B^{\mu\nu}-\frac{1}{4}W^I_{\mu\nu}W^{I\mu\nu}-\frac{1}{4}G^A_{\mu\nu}G^{A\mu\nu}\nonumber\\
	&+&(D_\mu H)^\dagger (D^\mu H)+\mu_H^2 H^\dagger H-\lambda (H^\dagger H)^2\nonumber\\
	&+&\sum_{f=q,u,d,\ell, e} i\bar{f}\slashed{D}f+iN_a \slashed{\partial} N_a -\frac{1}{2}M_{ab}\overline{N_a^c} N_b\nonumber\\
	&-&\left( \overline{q} Y_u\tilde{H} u+\overline{q} Y_d H d+\overline{\ell} Y_e He+\overline{\ell} Y_\nu\tilde{H} N+\rm h.c\right).
\end{eqnarray}
Here, we have defined the SM fields following the notation of \cite{SMEFTRGEs1,SMEFTRGEs2, SMEFTRGEs3}. We write the covariant derivatives as
\begin{equation}
	D_\mu=\partial_\mu+ ig'\mathcal{Y}B_\mu+ig\frac{\tau^I}{2}W^I_\mu+ig_s \frac{\lambda^A}{2} G^A\,,
\end{equation}
where $\mathcal{Y}$ represents the hypercharge, $I$ and $A$ are the adjoint indices of SU(2) and SU(3), and  $\tau^I$ and $\lambda^A$ are Pauli and Gell-Mann matrices, respectively. The electric charge is given by $Q = \mathcal{Y} + \tau^3/2$.  $\tilde{H}$ is defined as $\tilde{H}_I = \epsilon_{IJ} H^*_J$, where $\epsilon$ is the anti-symmetric SU(2) tensor with $\epsilon_{12} = -\epsilon_{21} = 1$. The matrices $Y_u$, $Y_d$, and $Y_e$ are $3\times3$ flavor matrices, while the neutrino Yukawa matrix $Y_\nu$ has dimensions $3\times n$.

\begin{table}[t!]

%
\mbox{}\\[-1.25cm]

\begin{adjustbox}{width=1.05\textwidth,center}
\begin{minipage}[t]{3cm}
\renewcommand{\arraystretch}{1.51}
\small
\begin{align*}
\begin{array}[t]{c|c}
	\multicolumn{2}{c}{\boldsymbol{\psi^2H^3 +{\rm h.c.}}} \\
\hline
	\mathcal{O}_{NH} & (\overline{\ell} \tilde{H} N)(H^\dagger H) \\
	\multicolumn{2}{c}{\boldsymbol{\psi^4: (\bar{R}R)(\bar{R}R)}} \\
\hline
	\mathcal{O}_{NN} & (\overline{N}\gamma^\mu N)(\overline{N}\gamma^\mu N)\\
 	\mathcal{O}_{Ne} & (\overline{N}\gamma^\mu N)(\overline{e}\gamma^\mu e)\\
	\mathcal{O}_{Nu} & (\overline{N}\gamma^\mu N)(\overline{u}\gamma^\mu u)\\
	\mathcal{O}_{Nd} & (\overline{N}\gamma^\mu N)(\overline{d}\gamma^\mu d)\\
	\mathcal{O}_{Nedu} & (\overline{N}\gamma^\mu e)(\overline{d}\gamma^\mu u)+{\rm h.c.}  \\
\end{array}
\end{align*}
\end{minipage}
\begin{minipage}[t]{3cm}
\renewcommand{\arraystretch}{1.51}
\small
\begin{align*}
\begin{array}[t]{c|c}
	\multicolumn{2}{c}{\boldsymbol{\psi^2XH  +{\rm h.c.}}} \\
\hline
	\mathcal{O}_{NB} & (\overline{\ell} \tilde{H} \sigma^{\mu\nu} N)B_{\mu\nu}\\
	\mathcal{O}_{NW} & (\overline{\ell} \tau^I\tilde{H} \sigma^{\mu\nu} N)W^I_{\mu\nu}\\
	\multicolumn{2}{c}{\boldsymbol{\psi^4: (\bar{R}R)(\bar{L}L)}} \\
\hline
	\mathcal{O}_{N\ell} & (\overline{N}\gamma^\mu N)(\overline{\ell}\gamma^\mu \ell)\\
	\mathcal{O}_{Nq} & (\overline{N}\gamma^\mu N)(\overline{q}\gamma^\mu q) \\
	\multicolumn{2}{c}{\boldsymbol{\psi^4: (\bar{R}L)(\bar{L}R) +{\rm h.c.}}} \\
\hline
	\mathcal{O}_{N\ell qu} & (\overline{N} \ell)(\overline{q} u) \\
\end{array}
\end{align*}
\end{minipage}
\begin{minipage}[t]{3cm}
\renewcommand{\arraystretch}{1.51}
\small
\begin{align*}
\begin{array}[t]{c|c}
	\multicolumn{2}{c}{\boldsymbol{\psi^2H^2D  +{\rm h.c.}}} \\
\hline
	\mathcal{O}_{HN} & (\overline{N}\gamma^\mu N)(H^\dagger i\overset{\leftrightarrow}{D}_\mu H) \\
	\mathcal{O}_{HNe} & (\overline{N}\gamma^\mu e)(\tilde{H}^\dagger i{D}_\mu H) \\
	\multicolumn{2}{c}{\boldsymbol{\psi^4: (\bar{L}R)(\bar{L}R)+{\rm h.c.}}} \\
\hline
	\mathcal{O}_{\ell N\ell e} & (\overline{\ell} N)\epsilon(\overline{\ell} e)\\
	\mathcal{O}_{\ell N qd} & (\overline{\ell} N)\epsilon(\overline{q} d)\\
	\mathcal{O}_{\ell d q N} & (\overline{\ell} d)\epsilon(\overline{q} N)  \\
\end{array}
\end{align*}
\end{minipage}
\end{adjustbox}
\setlength{\abovecaptionskip}{0.15cm}
\caption{Dimension six operators containing at least one sterile neutrino in the $\nu$SMEFT. We divide them into the same classes of SMEFT operators introduced in \cite{Grzadkowski:2010es}. The scalar operator $\mathcal{O}_{\ell d qN}$ can be replaced by the Fierz equivalent four-tensor $\mathcal{O}^{(3) \alpha a ij}_{\ell N q d}=-4\mathcal{O}^{\alpha a i j}_{\ell N qd}-8 \mathcal{O}^{\alpha j i a}_{\ell dq N}$.}\label{tab:nuSMEFTops}
\end{table}

The dimension six Lagrangian is given by
\begin{equation}
	\mathcal{L}^{(6)}=\frac{1}{\Lambda^2}\sum_{A,\,\xi}C^{\xi}_A \,\mathcal{O}^\xi_{A}\,,
\end{equation}
where {$C^{\xi}_A$ stands for the Wilson Coefficients (WCs),} $A$ runs over the physical operator basis, and $\xi$ sums over all possible flavor permutations for a given operator. We employ the dimension-six SMEFT basis from \cite{Buchmuller:1985jz, Grzadkowski:2010es} and organize them into the same operator classes. The new dimension-six operators of the $\nu$SMEFT are similarly categorized and summarized in Table~\ref{tab:nuSMEFTops}. To avoid double counting, we add the Hermitian conjugates only for non-self-conjugate operators, {\it i.e.}~those that do not differ from their Hermitian conjugates just for  flavour index permutations. 

\section{Computation of $\nu$SMEFT RGEs}
\label{sec:RGEs}
Our main goal is to compute the { RGEs} that describe the evolution within the $\nu$SMEFT from the new physics scale $\Lambda$ down to the sterile neutrino mass scale $M$ or the EW scale, the highest of both scales.
We focus on the mixing of the $\nu$SMEFT operators listed in Table~\ref{tab:nuSMEFTops} with both the SMEFT operators and themselves, and restrict our calculation to the anomalous dimension sub-matrix that contains at least one Yukawa insertion.
Once added to the other Yukawa terms\footnote{Our computation also includes mixed gauge-Yukawa and $\lambda$-Yukawa contributions, not only the {\it pure} Yukawa ones. We also calculate the self-running of $C_{NH}$ given by the Higgs self-coupling $\lambda$, although the corresponding anomalous dimension does not contain Yukawa couplings, and found agreement with~\cite{Chala:2020pbn}.} in the SMEFT~\cite{SMEFTRGEs1,SMEFTRGEs2,SMEFTRGEs3}, our results provide the complete Yukawa contributions to the $\nu$SMEFT RGEs, complementing the already available gauge coupling contributions~\cite{Datta:2020ocb}.

Schematically, the RGEs take the following form
\begin{equation}
	16\pi^2 \frac{d C_i}{d\log\mu}= \gamma_{ji}C_j\,,
 \label{eq:RGEs}
\end{equation}
where $C_i$ are the operator coefficients and $\gamma_{ji}$ is the anomalous dimension matrix, related to the divergent part of loop diagrams that mix the operators. 

To calculate the RGEs, we determine the divergent part of diagrams involving one $\nu$SMEFT operator dressed with Higgs field insertions. We work in 
$D=4-2\epsilon$ dimensions and renormalize using the modified minimal subtraction scheme ($\overline{\rm{MS}}$). We implement a model file in FeynArts~\cite{Hahn:2000kx} to include the $\nu$SMEFT dimension-six operators and generate the required diagrams. The $1/\epsilon$ poles are calculated both by hand and using FeynCalc~\cite{Mertig:1990an, Shtabovenko:2016sxi, Shtabovenko:2020gxv}.

We use the Equation of Motions (EOM) to project redundant counterterms into the physical operator basis specified in section~\ref{sec:nuSMEFT}. Whenever a loop diagram is renormalized by a redundant operator $\mathcal{O}_{\rm red}$,  which differs from a basis operator by an operator $\mathcal{O}_{\rm EOM}$ that vanishes when the classical EOM are satisfied, $ \mathcal{O}_{\rm red}=\mathcal{O}_{\rm phy}+\mathcal{O}_{\rm EOM}$, the counterterm
\begin{equation}
	\frac{A}{16\pi^2\epsilon}\mathcal{O}_{\rm red}\,,
\end{equation}
is equivalent to
\begin{equation}
	\frac{A}{16\pi^2\epsilon}\mathcal{O}_{\rm phy}\,,
\end{equation}
because EOM vanishing operators like $\mathcal{O}_{\rm EOM}$ lead to zero S-matrix elements, and cannot mix with the physical basis in the RGEs \cite{Simma:1993ky}.

The structure of the anomalous dimension sub-matrix we calculate is similar to the SMEFT Yukawa mixing involving up-type quarks. Whenever possible, we cross-checked our results against those in the SMEFT~\cite{SMEFTRGEs1,SMEFTRGEs2,SMEFTRGEs3} after performing the appropriate substitutions and found agreement. 
Notice however that this does not apply to the novel operator type introduced by adding sterile neutrinos, the two-lepton two-quark vector operator $\mathcal{O}_{Nedu}$ (see Table \ref{tab:nuSMEFTops}), which does not have an analogue in SMEFT. 
We also cross-checked with the partial results for the four-fermion operators~\cite{Datta:2021akg} and found agreement, up to some minor discrepancies, except for $\dot C_{\ell\ell}$.
Our RGE results are given in Appendix~\ref{App:RGEs}.

\section{Phenomenological implications: lepton flavor violation}
\label{sec:pheno}

As a phenomenological application of the RGEs derived in the previous section, we study now their implications for lepton flavor violation observables.
These exotic processes provide extremely sensitive probe for SMEFT operators and, as we will see, introducing sterile neutrinos with large Yukawa couplings can extend this sensitivity also to the new $\nu$SMEFT operators.

Large Yukawa couplings with the sterile neutrinos lead to sizable LFV rates already at the renormalizable level, stemming from their exchange in loop diagrams with charged lepton external legs (for a review, see {\it e.g.}~\cite{Abada:2018nio}). The dimension four LFV predictions, {\it i.e.}~not suppressed by $\Lambda$, are typically parametrised by the off-diagonal elements of the $\eta$ matrix~\cite{Fernandez-Martinez:2007iaa}
\begin{equation}
\eta=\frac12 v^2\,Y_\nu M^{-2} Y_\nu^\dagger\,.
\end{equation} 
The same matrix is in general responsible for non-unitary contributions to the lepton mixing matrix~\cite{Broncano:2002rw}, which can be probed by various lepton flavour conserving and violating observables~\cite{Blennow:2023mqx}. Although these effects significantly constrain the size of the LFV contributions from dimension four operators, it is still possible to have large flavor off-diagonal contributions from the new $\nu$SMEFT operators, which can then generate sizeable charged LFV transitions due to the operator mixing induced by the neutrino Yukawa insertions.

The key point here is that, in general, LFV transitions induced by $\nu$SMEFT operators are parametrised by a different combination of couplings, which can be large despite having suppressed $d=4$ contributions. 
In other words, the Yukawa texture can be such that $v^2(Y_\nu M^{-2} Y_\nu^\dagger)$ is diagonal\footnote{Notice that such configuration can be in agreement with neutrino oscillation data. For instance, in an inverse seesaw model, even diagonal $Y_\nu$ are possible if one chooses the proper $\mu$-term~\cite{Arganda:2014dta}.}, while for example $v^2/\Lambda^2(Y_\nu C_{Nf} Y_\nu^\dagger)$ is flavour changing.  In addition, some contributions feature only one insertion of a sterile Yukawa, which are generally less constrained.

\renewcommand{\arraystretch}{1.3}
\begin{table}[t!]
	\begin{center}
		\begin{tabular}{|l|c|c|}
			\hline
			Process & Current 90\%CL upper bound  & Upcoming Sensitivity    \\
			\hline
			$\mu\to e \gamma $ & $  3.1 \times 10^{-13}$ \cite{MEGII:2023ltw}
			& $  {\sim 6\times 10^{-14}}$ \cite{MEGII:2018kmf} \\
			$\mu\to \bar{e}ee $ & $  1.0  \times 10^{-12}$ \cite{SINDRUM:1987nra}
			& $ 10^{-14} \to 10^{-16}$ \cite{Mu3e:2020gyw} \\
			$\mu A \to e A$ &  $ 7 \times 10^{-13}$ \cite{SINDRUMII:2006dvw}
			&$\sim 10^{-16}$ \cite{COMET:2009qeh, Mu2e:2022ggl}   \\
			\hline
	\end{tabular}	\end{center} 
	\caption{Current bounds for some $\mu \to e$  processes, as well as the estimated reach  of upcoming   experiments. For $\mu\to e$ conversion in nuclei, A=Au(Al) for current (future) experiments.}
		\label{tab:bds}
\end{table}
\renewcommand{\arraystretch}{1}

In order to simplify our computations, we assume that the sterile neutrino mass scale to be of the order of the EW scale, so they are integrated out at the same time than the top, Higgs, $Z$ and $W$ bosons. 
This implies that our $\nu$SMEFT running, which takes the form
\begin{equation}
	C_A(m_W)=U_{BA}(\Lambda, m_W)\, C_B(\Lambda)\,,
\end{equation}
is valid from $\Lambda$ to the EW scale ($\sim m_W$), where we match the $\nu$SMEFT directly to the standard low-energy EFT without $N$~\cite{Jenkins:2017jig}.
If one considers instead the hypothesis $M > v$, the only difference would be applying the $\nu$SMEFT RGEs down to $M$, then integrating out the sterile neutrinos, and running with the SMEFT RGEs down to $m_W$.
Nevertheless, we consider that the $M\sim v$ hypothesis is enough to obtain a first feeling about the potential of LFV experiments to probe the new $\nu$SMEFT operators.

We focus on $\mu\to e$ transitions because they are the most constraining and expect a significant improvement of their sensitivities in the upcoming years (see Table~\ref{tab:bds}). To calculate the contributions of $\nu$SMEFT we solve the RGEs of Eq.~(\ref{eq:RGEs}) from $\Lambda$ to $m_W$, match onto the low-energy effective operator relevant for the $\mu\to e$ observables, and run down to the scale of the experiments in the effective theory with the QED and QCD invariant operators.
Neglecting the running of the SM couplings, the leading-log solution to the $\nu$SMEFT RGEs is\footnote{We also take into account the QCD running of the scalar and tensor operators, which results in a ${\rm few}\times 10 \%$ rescaling of their coefficients.  }
\begin{equation}
	C_A(m_W)=-\frac{\gamma_{BA}}{16\pi^2}\log\left(\frac{\Lambda}{m_W}\right) C_B(\Lambda)\,,\label{eq:leadinglog}
\end{equation}
where $\gamma$ is the anomalous dimension matrix. 

\begin{figure}[t!]
	\centering
\includegraphics[width=0.8\textwidth]{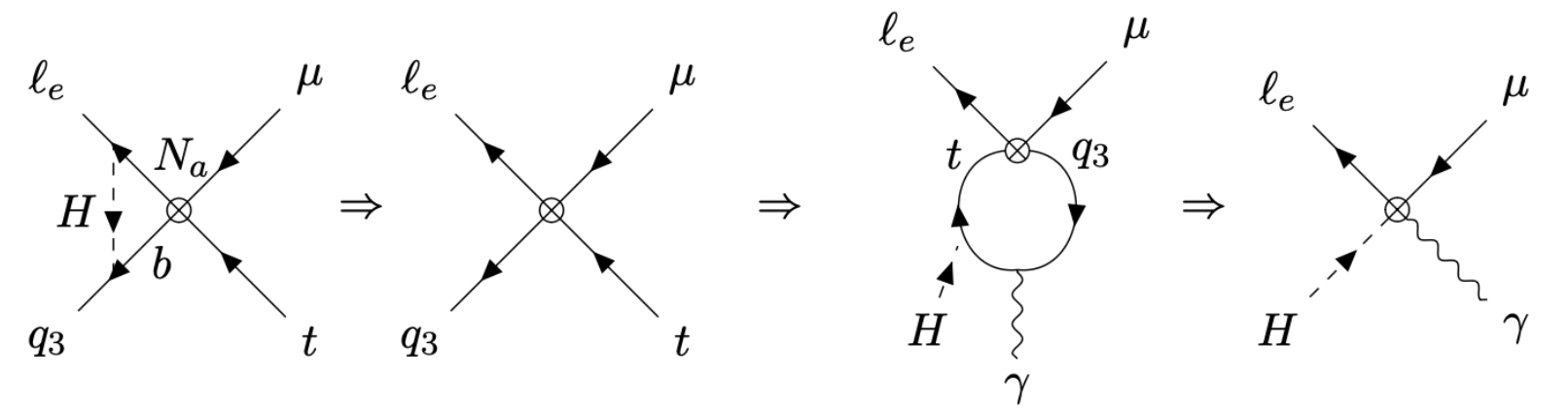}
\caption{Second order one-loop mixing of $C^{a \mu bt}_{Nedu}$ onto the $\mu\to e\gamma$ dipole. $C^{a \mu bt}_{Nedu}$ mixes with $C^{(3)e \mu tt}_{\ell e qu}$, which in turn can mix with $C^{e \mu}_{eB,W}$ with a top-Yukawa chiral enhancement. }\label{fig:2ndorderRGEs}
\end{figure}

Given the large number of operators that can contribute, we consider one operator at a time to find the sensitivities of $\mu\to e$ observables to $\nu$SMEFT operator coefficients. 
We summarise our results in Tables~\ref{tab:mutoe}$-$\ref{tab:mutoe2} for the operators to which $\mu\to e$ transitions are most sensitive.
In most cases, the leading-log approximation of Eq.~\eqref{eq:leadinglog} is sufficient to find the main contribution of each operator to $\mu\to e$ observables, however there are cases where second-order, $\mathcal O(\log^2/(16\pi^2)^2)$ effects are also relevant. For instance, the operator $C^{a\mu bt}_{Nedu}$ can mix with the tensor having top quarks $C^{(3)e\mu tt}_{\ell equ}$, which in turn can mix efficiently with the $\mu\to e\gamma$ dipole thanks to the chiral enhancement given by the large top Yukawa (see Figure~\ref{fig:2ndorderRGEs}). 
These second contributions are also included in Tables~\ref{tab:mutoe}$-$\ref{tab:mutoe2}.
The tables were obtained fixing the new physics scale $\Lambda$ to 1 TeV.

\begin{figure}[t!]
\begin{center}
\includegraphics[width=.45\textwidth]{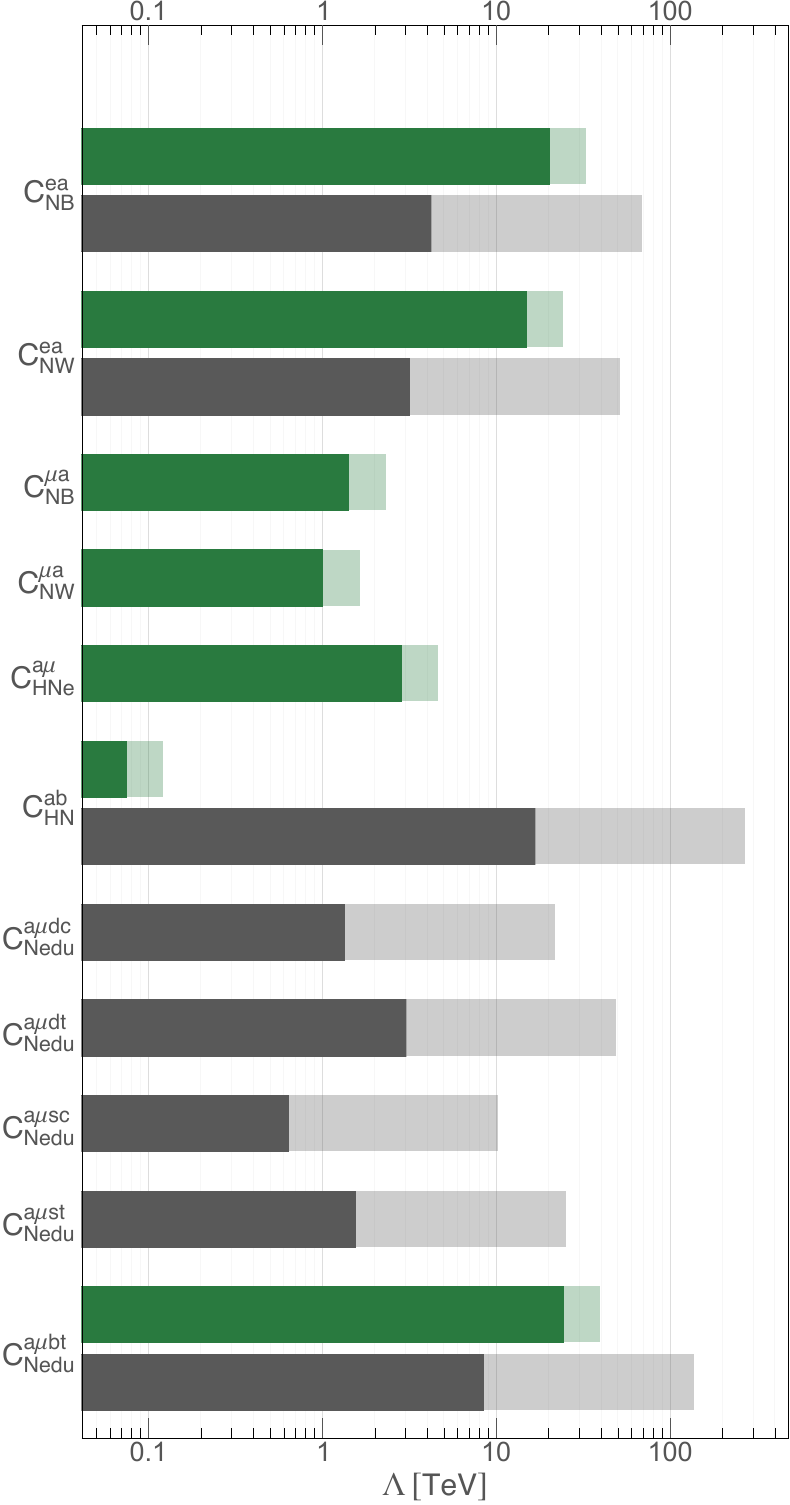}
\includegraphics[width=.45\textwidth]{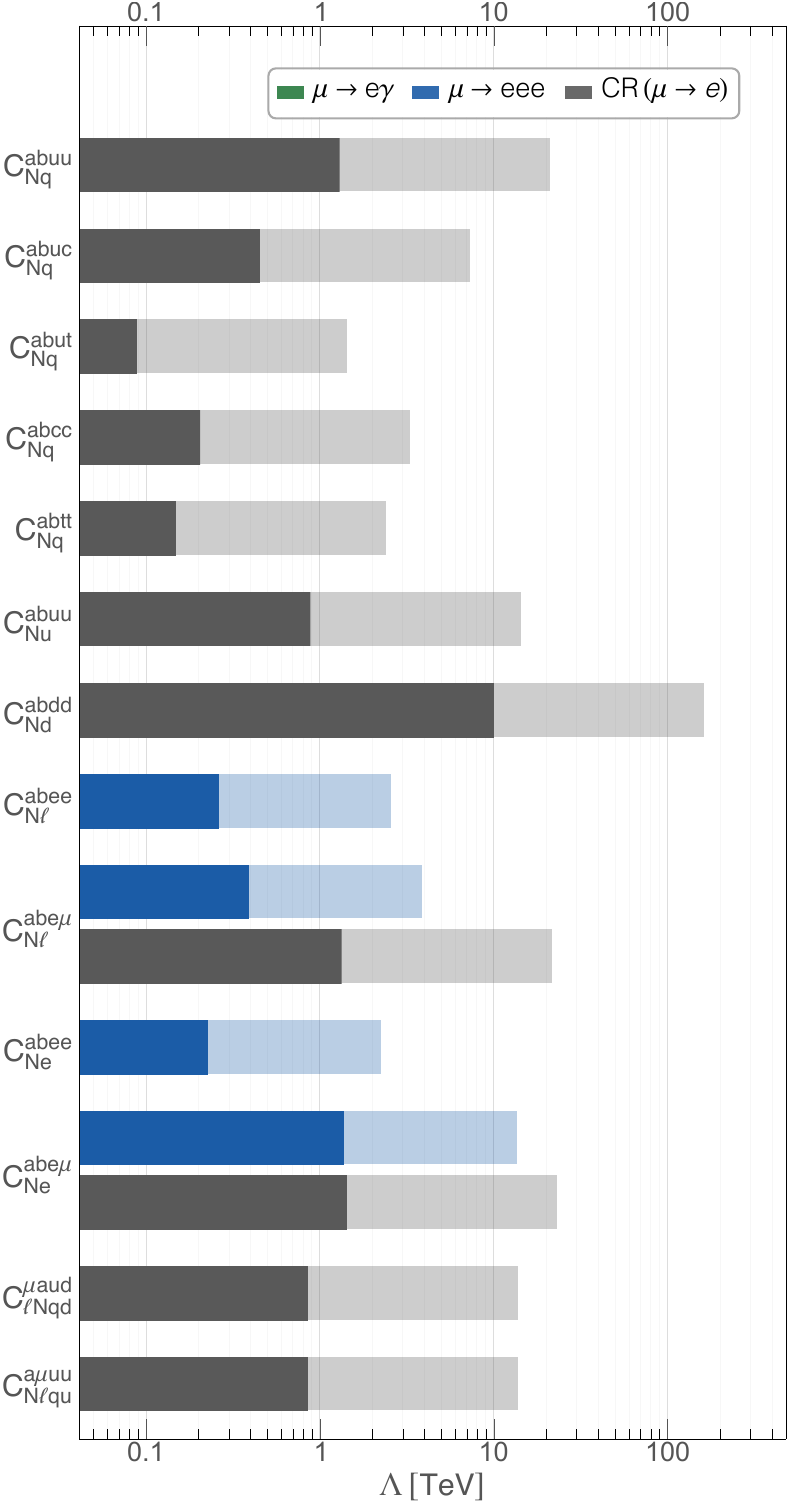}
\caption{New physics scale $\Lambda$ probed for selected $\nu$SMEFT LFV $\mu-e$ operators. Here we switch on each operator individually and consider diagonal neutrino Yukawa couplings of $10^{-2}$. Darker bars show current sensitivities, while lighter ones are for future sensitivities. }\label{fig:barplot}
\end{center}
\end{figure}

 Alternatively, we can fix the size of $Y_\nu$ and compute which scale we can probe for each observable, as shown in Fig.~\ref{fig:barplot}.
Here, and for concreteness, we have chosen a diagonal $Y_\nu$ matrix with entries of $10^{-2}$, which gives $\eta\sim \mathcal O(10^{-4})$, in agreement with the latest bounds~\cite{Blennow:2023mqx}.
Considering heavier masses would allow for larger Yukawa couplings, enhancing the anomalous couplings at the prize of (logarithmically) reducing the gap between $\Lambda$ and $M$.
All in all, we see that due to the impressive experimental sensitivities and the $Y_\nu$ RGE mixings, we are able to probe scales up to $\Lambda\sim10-100$~TeV for the new $\nu$SMEFT LFV operators.

\section{Conclusions}
\label{sec:conclusions}

The $\nu$SMEFT extends the general SMEFT to incorporate the well-motivated right-handed, sterile neutrinos, also known as Heavy Neutral Leptons. This interesting framework, however, is still less developed than the SMEFT itself. 
In particular, the role of the neutrino Yukawa coupling has been neglected despite the fact it can be large in symmetry protected low scale scenarios. 

In this work, we contributed to this task by computing the full Yukawa contributions to the one-loop anomalous dimensions. 
When combined with previous computations in the literature, our results provide the complete one-loop renormalization group evolution equations of the dimension 6 $\nu$SMEFT.

As an application of our results, we have explored the implications for the lepton flavor violating operators, focusing in particular on the $\mu-e$ sector. We showed that combining the relatively large neutrino Yukawa couplings expected in low scale seesaw models with the strong current experimental bounds on these rare processes, we can probe dimension 6 $\nu$SMEFT operators up to scales of $\Lambda\sim10-100$~TeV.

\medskip

\paragraph{Acknowledgments.}
The authors thank Sacha Davidson for very illuminating discussions during the first steps of this project, and Arsenii Titov for pointing missing references out and for very useful comments on the manuscript.  
This project has received support from the Spanish Research Agency (Agencia Estatal de Investigaci\'on) through the Grant IFT Centro de Excelencia Severo Ochoa No CEX2020-001007-S. The work of MA is supported by the Spanish AEI-MICINN PID2020-113334GB-I00/AEI/10.13039/501100011033.
XM acknowledges funding from the European Union’s Horizon Europe Programme under the Marie Skłodowska-Curie grant agreement no.~101066105-PheNUmenal. 

\begin{table}[h!]
\renewcommand{\arraystretch}{2}
\begin{tabular}{c|c|c}
Operator ($\Lambda = 1\ {\rm TeV})$ & 90\%CL upper limit &  Constraining observable \\
\hline
\multirow{2}{*}{$C^{e a}_{NB}$} & $ 2.5\times 10^{-5}\,\Big(\big| [Y_\nu]_{\mu a}\big|\Big)^{-1}  $ & $\mu\to e \gamma$ \\
 & $ 5.6\times 10^{-4}\,\Big(\big| [Y_\nu]_{\mu a}\big|\Big)^{-1}  $  & $\mu Au\to e Au$ \\
\hline
\hline
\multirow{2}{*}{$C^{e a}_{NW}$} & $ 4.6\times 10^{-5}\,\Big(\big| [Y_\nu]_{\mu a}\big|\Big)^{-1}  $ & $\mu\to e \gamma$ \\
 & $ 1\times 10^{-3}\,\Big(\big| [Y_\nu]_{\mu a}\big|\Big)^{-1}  $  & $\mu Au\to e Au$ \\
\hline
\hline
$C^{\mu a}_{NB}$ & $5\times 10^{-3}\,\Big(\big| [Y_\nu]_{e a}\big|\Big)^{-1} $  & $\mu\to e \gamma$ \\
\hline
\hline
$C^{\mu a}_{NW}$ & $1\times 10^{-2}\,\Big(\big| [Y_\nu]_{e a}\big|\Big)^{-1} $  & $\mu\to e \gamma$ \\
\hline
\hline
$C^{a\mu}_{HNe}$ & $1.5\times 10^{-3}\,\Big(\big| 2 [Y_\nu Y^\dagger_\nu Y_\nu]_{ea}-1.2 [Y_\nu]_{ea}\big|\Big)^{-1} $  & $\mu\to e \gamma$ \\
\hline
\hline
\multirow{2}{*}{$C^{ab}_{HN}$} & $  1.8\times 10^{-2}\,\Big(\big|[Y^*_\nu]_{\mu a} [Y_\nu]_{eb}\big|\Big)^{-1}$ & $\mu\to e \gamma$ \\
 & $ 3.6\times 10^{-5}\,\Big(\big|[Y^*_\nu]_{\mu a} [Y_\nu]_{eb}\big|\Big)^{-1}$ & $\mu Au\to e Au$ \\
\hline
\hline
$C^{a \mu d c}_{Nedu}$ & $5.6\times 10^{-3}\,\Big(\big| [Y_\nu]_{e a}\big|\Big)^{-1} $  & $\mu Au\to e Au$ \\
\hline
\hline
$C^{a \mu d t}_{Nedu}$ & $1.1\times 10^{-3}\,\Big(\big| [Y_\nu]_{e a}\big|\Big)^{-1} $  & $\mu Au\to e Au$ \\
\hline
\hline
$C^{a \mu s c}_{Nedu}$ & $2.5\times 10^{-2}\,\Big(\big| [Y_\nu]_{e a}\big|\Big)^{-1} $  & $\mu Au\to e Au$ \\
\hline
\hline
$C^{a \mu s t}_{Nedu}$ & $4.2\times 10^{-3}\,\Big(\big| [Y_\nu]_{e a}\big|\Big)^{-1} $  & $\mu Au\to e Au$ \\
\hline
\hline
\multirow{2}{*}{$C^{a\mu bt}_{Nedu}$} & $ 1.7\times 10^{-5}\,\Big(\big| [Y_\nu]_{e a}\big|\Big)^{-1}  $ & $\mu \to e \gamma$ \\
 & $ 1.4\times 10^{-4}\,\Big(\big| [Y_\nu]_{e a}\big|\Big)^{-1}   $  & $\mu Au\to e Au$ \\
\hline

\end{tabular}

\caption{Sensitivity of $\mu\to e$ processes to $\nu$SMEFT operators. We use up-type generation indices for the quark doublet to highlight that these limits are obtained in the $u-$basis, where $Y_u$ is diagonal. {These limits apply to the absolute values of the WCs.}}
\label{tab:mutoe}
\end{table}

\begin{table}[t!]
\renewcommand{\arraystretch}{2}
\begin{tabular}{c|c|c}
Operator ($\Lambda = 1\ {\rm TeV})$ & 90\%CL upper limit &  Constraining observable \\
\hline
$C^{ab uu}_{Nq}$ & $6\times 10^{-5}\,\Big(\big| [Y^*_\nu]_{\mu a} [Y_\nu]_{e b}\big|\Big)^{-1} $  & $\mu Au\to e Au$ \\
\hline
\hline
$C^{ab uc}_{Nq},\ C^{ab cu}_{Nq}$ & $5\times 10^{-4}\,\Big(\big| [Y^*_\nu]_{\mu a} [Y_\nu]_{e b}\big|\Big)^{-1} $  & $\mu Au\to e Au$ \\
\hline
\hline
$C^{ab ut}_{Nq},\ C^{ab tu}_{Nq}$ & $1.3\times 10^{-2}\,\Big(\big| [Y^*_\nu]_{\mu a} [Y_\nu]_{e b}\big|\Big)^{-1} $  & $\mu Au\to e Au$ \\
\hline
\hline
$C^{ab cc}_{Nq}$ & $2.4\times 10^{-3}\,\Big(\big| [Y^*_\nu]_{\mu a} [Y_\nu]_{e b}\big|\Big)^{-1} $  & $\mu Au\to e Au$ \\
\hline
\hline
$C^{ab tt}_{Nq}$ & $4.6\times 10^{-3}\,\Big(\big| [Y^*_\nu]_{\mu a} [Y_\nu]_{e b}\big|\Big)^{-1} $  & $\mu Au\to e Au$ \\
\hline
$C^{ab uu}_{Nu}$ & $1.3\times 10^{-4}\,\Big(\big| [Y^*_\nu]_{\mu a} [Y_\nu]_{e b}\big|\Big)^{-1} $  & $\mu Au\to e Au$ \\
\hline
\hline
$C^{ab dd}_{Nd}$ & $1\times 10^{-4}\,\Big(\big| [Y^*_\nu]_{\mu a} [Y_\nu]_{e b}\big|\Big)^{-1} $  & $\mu Au\to e Au$ \\
\hline
\hline
$C^{ab ee}_{N\ell}$ & $1.5\times 10^{-3}\,\Big(\big| [Y^*_\nu]_{\mu a} [Y_\nu]_{e b}\big|\Big)^{-1} $  & $\mu\to e \bar{e} e$ \\
\hline
\hline
\multirow{2}{*}{$C^{ab e \mu}_{N\ell}$} & $ 1.7\times 10^{-4}\,\Big(\big| [Y^*_\nu]_{\alpha a} [Y_\nu]_{\alpha b}\big|\Big)^{-1}  $ & $\mu Au\to e Au$ \\
 & $ 2\times 10^{-3}\,\Big(\big| [Y^*_\nu]_{\alpha a} [Y_\nu]_{\alpha b}\big|\Big)^{-1}   $  & $\mu \to e \bar{e} e$ \\
\hline
\hline
$C^{ab ee}_{N e}$ & $2\times 10^{-3}\,\Big(\big| [Y^*_\nu]_{\mu a} [Y_\nu]_{e b}\big|\Big)^{-1} $  & $\mu\to e \bar{e} e$ \\
\hline
\hline
\multirow{2}{*}{$C^{ab e \mu}_{N e}$} & $ 1.5\times 10^{-4}\,\Big(\big| [Y^*_\nu]_{\alpha a} [Y_\nu]_{\alpha b}\big|\Big)^{-1}  $ & $\mu Au\to e Au$ \\
 & $ 1.6\times 10^{-4}\,\Big(\big| [Y^*_\nu]_{\alpha a} [Y_\nu]_{\alpha b}\big|\Big)^{-1}   $  & $\mu \to e \bar{e} e$ \\
\hline
\hline
$C^{\mu a u d}_{\ell N qd}$ & $1.4\times 10^{-2}\,\Big(\big| [Y_\nu]_{e a}\big|\Big)^{-1} $  & $\mu Au\to e Au$ \\
\hline
\hline
$C^{a \mu u u}_{N\ell qu}$ & $1.4\times 10^{-2}\,\Big(\big| [Y_\nu]_{e a}\big|\Big)^{-1} $  & $\mu Au\to e Au$ \\
\hline
\hline
$C^{e a \beta\mu}_{\ell N \ell e}$ & $8\times 10^{-4}\,\Big(\big| [Y^\dagger_\nu Y_\nu Y^\dagger_\nu]_{a \beta} \big|\Big)^{-1} $  & $\mu\to e \gamma$ \\
\hline
\hline
$C^{\beta a e\mu}_{\ell N \ell e}$ & $3.7\times 10^{-4}\,\Big(\big| [Y^\dagger_\nu Y_\nu Y^\dagger_\nu]_{a \beta} \big|\Big)^{-1} $  & $\mu\to e \gamma$ \\
\hline
\hline
$C^{e a e\mu}_{\ell N \ell e}$ & $2.5\times 10^{-4}\,\Big(\big| [Y^\dagger_\nu Y_\nu Y^\dagger_\nu]_{a \mu} \big|\Big)^{-1} $  & $\mu\to e \gamma$ \\
\hline
\end{tabular}

\caption{Similar to Table \ref{tab:mutoe}. The lepton index $\beta$ can be $\mu$ or $\tau$, while $\alpha$ runs over all three generations (in $[Y^*_\nu]_{\alpha a} [Y_\nu]_{\alpha b}$ the sum is implicit).}
\label{tab:mutoe2}
\end{table}

\newpage
\appendix


\section{Anomalous Dimensions}\label{App:RGEs}
In this appendix, we write the renormalization group equations. We use the short-hand notation
\begin{equation}
    \dot{C}\equiv (16\pi^2)\frac{d C}{d\log \mu}\,,
\end{equation}
where $\mu$ is the renormalization scale. 
We collect only those terms containing Yukawa contributions that are not already present in the SMEFT anomalous dimensions~\cite{SMEFTRGEs1, SMEFTRGEs2, SMEFTRGEs3}. This means that, in order to obtain the complete Yukawa contributions in the $\nu$SMEFT, the new terms in this appendix must be added to the SMEFT ones. 

Furthermore, there are some definitions in Refs.~\cite{SMEFTRGEs1, SMEFTRGEs2, SMEFTRGEs3} that are also affected by the neutrino Yukawa coupling. 
On the one hand, the wavefunction renormalization contributions proportional to Yukawa couplings obtain a new contribution\footnote{Notice that our definition of Yukawa coupling relates to that of Refs.~\cite{SMEFTRGEs1, SMEFTRGEs2, SMEFTRGEs3} by an hermitian conjugate.}:
\begin{equation}
    \gamma_\ell^{(Y)} = \frac12 [Y_e^{}Y_e^\dagger + Y_\nu^{}Y_\nu^\dagger] \,, 
    \quad
    \gamma_H^{(Y)} = \Tr\big[ N_cY_u^{}Y_u^\dagger + N_cY_d^{\phantom\dagger}Y_d^\dagger + Y_e^{}Y_e^\dagger + Y_\nu^{}Y_\nu^\dagger\big]\,,
\end{equation}
as well as a new one associated to $N$:
\begin{equation}
    \gamma_N^{(Y)} = \frac12 [Y_\nu^{}Y_\nu^\dagger] \,.
\end{equation}
On the other hand, the $\xi$ and $\eta$ parameters must be extended as
\begin{equation}\label{eq:new_eta_xi}
    \eta_{i} = \eta_{i}^{\rm SMEFT} + \eta_i^\nu\,,
    \quad
    \xi_f = \xi_f^{\rm SMEFT} + \xi_f^\nu\,,
\end{equation}
where the SMEFT contributions are as in Refs.~\cite{SMEFTRGEs1, SMEFTRGEs2, SMEFTRGEs3}, and the new terms due to the neutrino Yukawa coupling read:
\begin{align}
    \eta_1^\nu & = \frac{1}{2} C_{NH}^{\alpha a} [Y^\dagger_\nu]_{a\alpha} 
    +\frac{1}{2} C_{NH}^{*\alpha a} [Y^{}_\nu]_{\alpha a}\,,
    \\
    \eta_2^\nu & = C_{HNe}^{a\alpha} [Y_e^\dagger Y_\nu^{}]_{\alpha a} + C_{HNe}^{*a\alpha} [Y_\nu^\dagger Y_e^{}]_{a\alpha} 
    -2C_{H\ell}^{(3)\,\alpha\beta} [Y_\nu^{} Y_\nu^\dagger]_{\beta\alpha}\,,
    \\
    \eta_3^\nu & = C_{HN}^{ab} [Y_\nu^\dagger Y_\nu^{}]_{ba}
    -C_{HNe}^{a\alpha} [Y_e^\dagger Y_\nu^{}]_{\alpha a} - C_{HNe}^{*a\alpha} [Y_\nu^\dagger Y_e^{}]_{a\alpha}
    -\big[C_{H\ell}^{(1)} -3C_{H\ell}^{(3)}\big]^{\alpha\beta} [Y_\nu^{} Y_\nu^\dagger]_{\beta\alpha}\,,
    \\
    \eta_4^\nu & =4C_{HN}^{ab} [Y_\nu^\dagger Y_\nu^{}]_{ba}
    +2C_{HNe}^{a\alpha} [Y_e^\dagger Y_\nu^{}]_{\alpha a} +2 C_{HNe}^{*a\alpha} [Y_\nu^\dagger Y_e^{}]_{a\alpha}
    -4C_{H\ell}^{(1)\,\alpha\beta} [Y_\nu^{} Y_\nu^\dagger]_{\beta\alpha}\,, \\
    \eta_5^\nu & = \frac{i}{2} C_{NH}^{\alpha a} [Y^\dagger_\nu]_{a\alpha} 
    -\frac{i}{2} C_{NH}^{*\alpha a} [Y^{}_\nu]_{\alpha a}\,, 
    \\
    %
    %
    [\xi^\nu_u]_{ij} & = -C^{a\alpha ij}_{N\ell qu} [Y_\nu]_{\alpha a}
    \\
    [\xi^\nu_d]_{ij} & = -\big(C^{\alpha a ij}_{\ell N qd}-1/2 C^{\alpha ji a}_{\ell d qN}\big)[Y^\dagger_\nu]_{a\alpha}\,,
    \\
    [\xi^\nu_e]_{\alpha\beta} & = -\big(C^{\rho a \alpha {\beta}}_{\ell N \ell e}+1/2 C^{\alpha a \rho {\beta}}_{\ell N \ell e}\big)[Y^\dagger_\nu]_{a\rho}\,.
\end{align}
And, finally, the new coefficient in $\nu$SMEFT:
\begin{align}
    [\xi_N]_{\alpha b} & = 2 C^{c b\alpha \rho }_{N \ell}[Y_\nu]_{\rho c}
    -\big(C^{\alpha b \rho \delta}_{\ell N \ell e}+\frac12 C^{\rho b \alpha\delta}_{\ell N \ell e}\big)[Y^\dagger_e]_{\delta \rho} 
    -N_c \big(C^{\alpha b kl}_{\ell N qd}-\frac12 C^{\alpha l k b}_{\ell d q N}\big)[Y^\dagger_d]_{lk} 
    \nonumber\\
    &-N_c C^{*b\alpha kl}_{N\ell qu}[Y_{u}]_{kl}\,.
\end{align}

Through the results of this appendix, we will not write explicitly the new contributions to the anomalous dimensions of the SMEFT WCs that are already taken into account by the modifications of $\gamma_{\{\ell,H\}}$, $\eta_i$ or $\xi_i$. The only exception will be in the mixed Yukawa-$\lambda$ contributions, since Ref.~\cite{SMEFTRGEs1} did not use $\xi_i$ to express the terms originated from the equation of motion. Conversely, if the SMEFT anomalous dimensions are expressed in terms of the $\xi_i$, our new terms proportional to $\lambda \xi_i^\nu$ will be already taken into account by the redefinition in Eq.~\eqref{eq:new_eta_xi} and needed to be removed from the following equations.

\subsection[$H^6$]{$\boldsymbol{H^6}$}

\begin{equation}
    {
    \dot{C}_H=-4 C_{NH}^{\alpha a} [Y^\dagger_\nu Y^{}_\nu Y^\dagger_\nu]_{a\alpha} -4 C_{NH}^{*\alpha a} [Y^{}_\nu Y^\dagger_\nu Y^{}_\nu]_{\alpha a}\,.
    }
\end{equation}

\subsection[$X^2H^2$]{$\boldsymbol{X^2H^2}$}
\begin{align}
    \dot{C}_{HW} & =-g C_{NW}^{\alpha a} [Y^\dagger_\nu]_{a\alpha} -g C^{*\alpha a}_{NW} [Y_\nu]_{\alpha a}\,, 
    \\
    \dot{C}_{H\tilde{W}} & = ~ig C_{NW}^{\alpha a} [Y^\dagger_\nu]_{a\alpha} -ig C^{*\alpha a}_{NW} [Y_\nu]_{\alpha a}\,, 
    \\
    \dot{C}_{HB} & =~g' C_{NB}^{\alpha a} [Y^\dagger_\nu]_{a\alpha} +g' C^{*\alpha a}_{NB} [Y_\nu]_{\alpha a}\,,
    \\
    \dot{C}_{H\tilde B} & =-ig' C_{NB}^{\alpha a} [Y^\dagger_\nu]_{a\alpha} +ig' C^{*\alpha a}_{NB} [Y_\nu]_{\alpha a}\,,
    \\
    \dot{C}_{HWB} & =~g C_{NW}^{\alpha a} [Y^\dagger_\nu]_{a\alpha} +g C^{*\alpha a}_{NW} [Y_\nu]_{\alpha a}
    -g' C_{NB}^{\alpha a} [Y^\dagger_\nu]_{a\alpha} -g' C^{*\alpha a}_{NW} [Y_\nu]_{\alpha a}\,,
    \\
    \dot{C}_{H\tilde WB} & =-ig C_{NW}^{\alpha a} [Y^\dagger_\nu]_{a\alpha} +ig C^{*\alpha a}_{NW} [Y_\nu]_{\alpha a}
    +ig' C_{NB}^{\alpha a} [Y^\dagger_\nu]_{a\alpha} -ig' C^{*\alpha a}_{NW} [Y_\nu]_{\alpha a}\,.
\end{align}

\subsection[$\psi^2 H^3$]{$\boldsymbol{\psi^2 H^3}$}

\begin{align}
    \dot{C}^{\alpha a }_{NH} &=  
        2(\eta_1+\eta_2-i\eta_5)[Y_\nu]_{\alpha a}
        + [Y^{}_\nu Y^\dagger_\nu Y^{}_\nu]_{\alpha a} (C_{HD}-6C_{H\square})
        -2C^{(1)\,\alpha\rho}_{H\ell} [Y^{}_\nu Y^\dagger_\nu Y^{}_\nu]_{\rho a}
        \nonumber\\
        &
        +6C^{(3)\,\alpha\rho}_{H\ell}[Y^{}_e Y^\dagger_e Y^{}_\nu]_{\rho a}
        +[Y^{}_\nu Y^\dagger_\nu Y^{}_\nu]_{\alpha b}C^{ba}_{HN}
        -2[Y^{}_e Y^\dagger_e Y^{}_e]_{\alpha \beta} C^{*a\beta}_{HNe}
        +8[Y^{}_\nu Y^\dagger_\nu Y^{}_\nu]_{\beta b }C^{ ba\alpha\beta}_{N\ell}
        \nonumber\\
        &
        -4[Y^\dagger_e Y^{}_e Y^\dagger_e]_{\rho \sigma}(C^{\alpha a \sigma \rho}_{\ell N\ell e}+1/2C^{\sigma a \alpha \rho}_{\ell N\ell e})
        -4{N_c}[Y^\dagger_d Y^{}_d Y^\dagger_d]_{ji}(C^{\alpha a ij}_{\ell N qd}-1/2C^{\alpha j i a}_{\ell d qN})
        \nonumber\\
        &
        +4[Y^\dagger_\nu Y^{}_\nu]_{ba}C^{\alpha b}_{NH}
        +5[Y^{}_\nu Y^\dagger_\nu]_{\alpha \sigma} C^{\sigma a}_{NH}
        -2 [Y_e]_{\alpha\rho} C_{eH}^{*\rho\sigma} [Y_\nu]_{\sigma a}
        - C_{eH}^{\alpha\rho} [Y^\dagger_e Y^{}_\nu]_{\rho a}
        -2[Y^{}_e Y^\dagger_e]_{\alpha \sigma} C^{\sigma a}_{NH}
        \nonumber\\
        &
        -4 N_c[Y^{}_u Y^\dagger_u Y^{}_u]_{ij} C_{N\ell qu}^{*a \alpha ij}
        +3\gamma_H^{(Y)} C_{NH}^{\alpha a} 
        + [\gamma_\ell^{(Y)}]_{\alpha\beta} C_{NH}^{\beta a} 
        + C_{NH}^{\alpha b}[\gamma_N^{(Y)}]_{ba}
        \nonumber \\
        &
        +[Y_\nu]_{\alpha a}\left[\frac{10}{3} g^2 C_{H\square}-\frac{3}{2}\left(g^2-g^{\prime 2}\right) C_{HD}\right]
        +\frac{4}{3}g^2 [Y_\nu]_{\alpha a}\left(C^{(3)\beta \beta}_{H\ell}+N_c C^{(3)ii}_{Hq}\right)
        \nonumber \\
        &
        +3[Y_\nu]_{\alpha a}\Big(3g^2(C_{HW}+iC_{H\tilde{W}})+g^{\prime 2}(C_{HB}+iC_{H\tilde{B}})+ g g'(C_{HWB}+iC_{H\tilde{W}B})\Big)
        \nonumber \\
        &
        -12 g [Y^{}_e Y^\dagger_e]_{\alpha \rho} C^{\rho a}_{NW}
        -6g C_{eW}^{\alpha\rho} [Y^\dagger_e Y^{}_\nu]_{\rho a}
        -3 \big(g C_{NW}^{\alpha b} {-} g'C_{NB}^{\alpha b}\big)[Y^\dagger_\nu Y^{}_\nu]_{b a}
        \nonumber \\
        &
        + 3 [Y^{}_\nu Y^\dagger_\nu]_{\alpha\rho} \big(g C_{NW}^{\rho a}+g' C_{NB}^{\rho a}\big)
        {-3(g^2+g'^2) [Y_\nu]_{\alpha b} C_{HN}^{b a}}
        +3g^2 [Y_e]_{\alpha \sigma} C^{*a\sigma}_{HNe} 
        &
        \nonumber \\
        & 
        +4\lambda\Big\{
        { 6 C^{\alpha a}_{NH}}
        - C^{(1)\alpha \beta}_{H\ell}[Y_\nu]_{\beta a}
        +3 C^{(3)\alpha \beta}_{H\ell}[Y_\nu]_{\beta a}
        +  [Y_\nu]_{\alpha b} C^{ba}_{HN}
        - [Y_e]_{\alpha \beta} C^{* a \beta}_{HNe}
        &
        \nonumber \\
        &-[Y_\nu]_{\alpha a}(C_{H\square}-\frac12 C_{HD})
        -  [\xi_N]_{\alpha a} \Big\}\,,
        \\
    \nonumber\\
    \dot{C}^{\alpha \beta }_{eH} &=  
        6 C^{(3)\alpha \rho}_{H\ell}[Y^{}_\nu Y^\dagger_\nu Y^{}_e]_{\rho \beta}
        -2[Y^{}_\nu Y_\nu^\dagger Y^{}_\nu]_{\alpha a} C^{a \beta}_{HNe}
        -2(2C^{\rho a \alpha\beta}_{\ell N \ell e}+C^{\alpha a \rho\beta}_{\ell N \ell e})[Y^\dagger_\nu Y^{}_\nu Y^\dagger_\nu]_{a \rho}
        \nonumber \\
        &
        -2[Y_\nu]_{\alpha a}C^{*\rho a}_{NH} [Y_e]_{\rho\beta}
        -C^{\alpha a}_{NH}[Y_\nu^\dagger Y^{}_e]_{a\beta}
        -2[Y_\nu Y^\dagger_\nu]_{\alpha \rho} C^{\rho \beta}_{eH}
        \nonumber \\
        &
        -12 g[Y_\nu Y^\dagger_\nu]_{\alpha\rho} C^{\rho \beta}_{eW}
        -6g C^{\alpha a}_{NW} [Y^\dagger_\nu Y_e]_{a\beta}
        +3g^2[Y_\nu]_{\alpha a} C^{a \beta}_{HNe}
        &
        \nonumber \\
        & 
        -4\lambda \Big\{[Y_\nu]_{\alpha a}C^{a \beta}_{HNe}
        +[\xi^\nu_e]_{\alpha \beta}\Big\}\,,
        \\
    \nonumber\\
    \dot{C}^{ij}_{uH} &=  
        - 4 [Y^{}_\nu Y^\dagger_\nu Y^{}_\nu]_{\alpha a} C_{N\ell qu}^{a\alpha ij}
        -4\lambda [\xi^\nu_u]_{ij} \,.
        \\
    \nonumber\\
    \dot{C}^{ij}_{dH} &=  
        -4 \big(C^{\alpha a ij}_{\ell N qd}-1/2 C^{\alpha ji a}_{\ell d qN}\big) [Y^{\dagger}_\nu Y_\nu Y^{\dagger}_\nu]_{a\alpha}
        -4\lambda [\xi^\nu_d]_{ij} \,.
\end{align}

\subsection[$\psi^2 XH$]{$\boldsymbol{\psi^2 XH}$}

\begin{align}
    \dot{C}^{\alpha a }_{NB} &=  
        2[Y^{}_\nu Y_\nu^\dagger-Y^{}_e Y_e^\dagger]_{\alpha \rho}C^{\rho a}_{NB}
        +C^{\alpha b}_{NB}[Y^\dagger_\nu Y^{}_\nu]_{ba}
        -C^{\alpha \rho}_{eB} [Y^\dagger_e Y^{}_\nu]_{\rho a}
        \nonumber \\
        &
        +\gamma_H^{(Y)} C_{NB}^{\alpha a}
        + [\gamma_\ell^{(Y)}]_{\alpha\beta} C_{NB}^{\beta a} 
        + C_{NB}^{\alpha b}[\gamma_N^{(Y)}]_{ba}
        \nonumber \\
        &
        +{\frac{1}{12}}g'N_c[Y^\dagger_d]_{ij}C^{\alpha j ia}_{\ell d qN}
        +\frac{3}{4}g'[Y^\dagger_e]_{\rho\sigma }C^{ \sigma a\alpha \rho} _{\ell N \ell e}
        \nonumber \\
        &
        +[Y_\nu]_{\alpha a}\Big(g'(C_{HB}+iC_{H\tilde{B}})+\frac{3}2g(C_{HWB}+iC_{H\tilde{W}B})\Big)\,,
        \\
    \nonumber\\
    \dot{C}^{\alpha a }_{NW} &=  
        2[Y^{}_e Y_e^\dagger]_{\alpha \rho}C^{\rho a}_{NW}
        +C^{\alpha b}_{NW}[Y^\dagger_\nu Y^{}_\nu]_{ba}
        -C^{\alpha \rho}_{eW} [Y^\dagger_e Y^{}_\nu]_{\rho a}
        \nonumber \\
        &
        +\gamma_H^{(Y)} C_{NW}^{\alpha a}
        + [\gamma_\ell^{(Y)}]_{\alpha\beta} C_{NW}^{\beta a} 
        + C_{NW}^{\alpha b}[\gamma_N^{(Y)}]_{ba}
        \nonumber \\
        &
        +\frac{1}{4}gN_c[Y^\dagger_d]_{ij} C^{\alpha j ia}_{\ell d qN}
        +\frac{1}{4}g[Y^\dagger_e]_{ \rho \sigma }C^{ \sigma a\alpha \rho} _{\ell N \ell e}
        \nonumber \\
        &
        -[Y_\nu]_{\alpha a}\Big(g(C_{HW}+iC_{H\tilde{W}})+\frac{1}2g'(C_{HWB}+iC_{H\tilde{W}B})\Big)\,,
        \\
    \nonumber\\
    \dot{C}^{\alpha \beta }_{eB} &=  
        -C^{\alpha a}_{NB}[Y^\dagger_\nu Y_e]_{a\beta}
        -2[Y_\nu Y^\dagger_\nu]_{\alpha\rho}C^{\rho \beta}_{eB}
        {
        + \frac14g' C_{\ell N\ell e}^{\alpha b\sigma \beta} [Y^\dagger_\nu]_{b\sigma}\,,
        }
        \\
    \nonumber\\
    \dot{C}^{\alpha \beta }_{eW} &=  
        -C^{\alpha a}_{NW}[Y^\dagger_\nu Y_e]_{a\beta}
        +2[Y_\nu Y^\dagger_\nu ]_{\alpha \rho}C^{\rho \beta}_{eW}
        {
        + \frac14g C_{\ell N\ell e}^{\alpha b\sigma \beta} [Y^\dagger_\nu]_{b\sigma}\,.
        }
\end{align}

\subsection[$\psi^2 H^2D$]{$\boldsymbol{\psi^2 H^2D}$}
\begin{align}
    \dot{C}^{ab}_{HN} &=  
        -[Y^\dagger_\nu Y^{}_\nu]_{ab}(C_{H\square}+C_{HD})
        -2[Y^\dagger_\nu]_{a\rho }C^{(1)\rho \sigma }_{H\ell}[Y_\nu]_{\sigma b}+
        3C^{ac}_{HN}[Y^\dagger_\nu Y^{}_\nu]_{cb}
        +3[Y^\dagger_\nu Y^{}_\nu]_{ac} C^{cb}_{HN}
        \nonumber \\
        &
        +[Y^\dagger_\nu Y^{}_e]_{a\rho} C^{* b \rho}_{HNe} +C^{a\rho}_{HNe}[Y^\dagger_e Y^{}_\nu ]_{\rho b}
        -4 [Y^\dagger_{\nu} Y^{}_\nu]_{dc}{(C^{abcd}_{NN}+C^{adcb}_{NN})}
        \nonumber\\
        &
        -2N_c [Y^\dagger_u Y^{}_u]_{kj} C^{abjk}_{Nu} 
        -2N_c [Y^{}_d Y^\dagger_d]_{kj} C^{abjk}_{Nq}
        +2N_c[Y^{}_u Y_u^\dagger]_{kj}C^{abjk}_{Nq}
        +2N_c [Y^\dagger_d Y^{}_d]_{kj} C^{abjk}_{Nd}
        \nonumber\\
        &
        +2 [Y^\dagger_e Y^{}_e]_{\rho \beta} C^{ab\beta \rho}_{Ne}
        +2 [Y^{}_\nu Y^\dagger_\nu]_{\beta \alpha} C^{ab\alpha \beta}_{N\ell}
        -2[Y^{}_e Y^\dagger_e]_{\rho \beta}C^{ab\beta \rho}_{N\ell}
        \nonumber\\
        &
        +2\gamma_H^{(Y)} C^{ab}_{HN}
        + [\gamma_N^{(Y)}]_{ac} C_{HN}^{c b} 
        +  C_{HN}^{a c} [\gamma_N^{(Y)}]_{cb} \,,
        \\
    \nonumber\\
    \dot{C}^{a\alpha}_{HNe} &=  
        [Y^\dagger_\nu Y^{}_e]_{a \alpha}(2 C_{H\square}-C_{HD})
        -2[Y^\dagger_\nu Y^{}_e]_{a\rho} C^{\rho \alpha}_{He}
        +2C^{ab}_{HN} [Y^\dagger_\nu Y^{}_e]_{b \alpha}
        +4[Y^\dagger_\nu Y^{}_e]_{b \sigma}C^{a b \sigma \alpha }_{Ne}
        \nonumber \\
        &
        +2[Y^\dagger_\nu Y^{}_\nu]_{ab} C^{b\alpha}_{HNe}
        + 2C^{a \beta}_{HNe}[Y^\dagger_e Y^{}_e]_{\beta \alpha}
        +4N_c[Y^\dagger_uY^{}_d]_{ik}C^{a\alpha ki}_{Nedu}
        \nonumber \\
        &
        +2\gamma_H^{(Y)} C^{a\alpha}_{HNe} 
        +[\gamma_N^{(Y)}]_{ac} C_{HNe}^{c \alpha} 
        + C_{HNe}^{a \beta} [\gamma_e^{(Y)}]_{\beta \alpha}\,,
        \\
    \nonumber\\
    \dot{C}^{\alpha\beta}_{He}& =  
        -[Y_e^\dagger Y^{}_\nu]_{\alpha a}C^{a \beta}_{HNe}
        -C^{*a\alpha}_{HNe}[Y^\dagger_\nu Y^{}_e  ]_{a\beta}
        -2[Y^\dagger_\nu Y^{}_\nu]_{ba}C^{ab\alpha\beta}_{Ne}
        +2[Y^{}_\nu Y^\dagger_\nu]_{\rho \sigma}C^{\sigma \rho \alpha \beta}_{\ell e}\,,
        \\
    \nonumber\\
    \dot{C}^{(1)\alpha\beta}_{H\ell} &=  
        \frac{1}{2}[Y^{}_\nu Y^\dagger_\nu]_{\alpha \beta}(C_{H\square} + C_{HD})
        -[Y^{}_\nu]_{\alpha a}[Y^\dagger_{\nu}]_{b\beta} C^{ab}_{HN} 
        -2[Y^\dagger_\nu Y^{}_\nu]_{ba}C^{ab\alpha\beta}_{N\ell}
        \nonumber \\
        &
        +\frac{3}{2}\Big([Y^{}_\nu Y^\dagger_\nu]_{\alpha \rho} C^{(1)\rho \beta}_{H\ell}
        +C^{(1)\alpha \rho}_{H\ell} [Y^{}_\nu Y^\dagger_\nu ]_{\rho \beta}\Big)
        -\frac{9}{2}\Big([Y^{}_\nu Y^\dagger_\nu]_{\alpha \rho} C^{(3)\rho \beta}_{H\ell}
        +C^{(3)\alpha \rho}_{H\ell} [Y^{}_\nu Y^\dagger_\nu ]_{\rho \beta}\Big)
        \nonumber\\
        &
        +2[Y^{}_\nu Y^\dagger_\nu]_{\rho \sigma}\Big({2C^{\alpha \beta \sigma \rho}_{\ell \ell}+C^{\alpha \rho \sigma \beta}_{\ell \ell}}\Big)\,,
        \\
    \nonumber\\
    \dot{C}^{(3)\alpha\beta}_{H\ell} &=  
        -\frac{1}{2}[Y^{}_\nu Y^\dagger_\nu]_{\alpha \beta} C_{H\square}
        -\frac{3}{2}\Big([Y^{}_\nu Y^\dagger_\nu]_{\alpha \rho} C^{(1)\rho \beta}_{H\ell}
        +C^{(1)\alpha \rho}_{H\ell} [Y^{}_\nu Y^\dagger_\nu ]_{\rho \beta}\Big)
        \nonumber\\
        &
        +\frac{1}{2}\Big([Y^{}_\nu Y^\dagger_\nu]_{\alpha \rho} C^{(3)\rho \beta}_{H\ell}
        +C^{(3)\alpha \rho}_{H\ell} [Y^{}_\nu Y^\dagger_\nu ]_{\rho \beta}\Big)
        -2[Y^{}_\nu Y^\dagger_\nu]_{\rho \sigma}{C^{\alpha \rho \sigma \beta}_{\ell \ell}}\,,
        \\
    \nonumber\\
    \dot{C}^{(1)ij}_{Hq} &=  
        -2[Y^\dagger_\nu Y^{}_\nu]_{ba}C^{abij}_{Nq}
        +2[Y^{}_\nu Y^\dagger_\nu]_{\sigma \rho}C^{(1)\rho \sigma ij}_{\ell q}\,,
        \\
    \nonumber\\
    \dot{C}^{(3)ij}_{Hq} &=  
        {
        -}2[Y^{}_\nu Y^\dagger_\nu]_{\sigma \rho}C^{(3)\rho \sigma ij}_{\ell q}\,,
        \\
    \nonumber\\
    \dot{C}^{ij}_{Hud} &=  
        -2[Y^\dagger_\nu Y^{}_e]_{a\alpha}C^{*a\alpha ji}_{Nedu}\,,
        \\
    \nonumber\\
    \dot{C}^{ij}_{Hu} &=  
        -2[Y^\dagger_\nu Y^{}_\nu]_{ba}C^{abij}_{Nu}{+2[Y_\nu Y^\dagger_\nu]_{\sigma \rho}C^{\rho \sigma ij}_{\ell u}}\,,
        \\
    \nonumber\\
    \dot{C}^{ij}_{Hd} &=  
        -2[Y^\dagger_\nu Y^{}_\nu]_{ba}C^{abij}_{Nd}{+2[Y_\nu Y^\dagger_\nu]_{\sigma \rho}C^{\rho \sigma ij}_{\ell d}}\,.
\end{align}

\subsection[$\psi^4$]{$\boldsymbol{\psi^4}$}

\subsubsection[$(\bar{R}R)(\bar{R}R)$]{$\boldsymbol{(\bar{R}R)(\bar{R}R)}$}
\begin{align}
    \dot{C}^{abcd}_{NN} &=  
        -[Y^\dagger_\nu Y^{}_\nu]_{cd}C^{ab}_{HN}
        -[Y^\dagger_\nu Y^{}_\nu]_{ab}C^{cd}_{HN}
        -[Y^\dagger_\nu]_{c \rho}[Y^{}_\nu]_{\sigma d}C^{ab\rho \sigma }_{N \ell }
        -[Y^\dagger_\nu]_{a \rho}[Y^{}_\nu]_{\sigma b}C^{cd\rho \sigma }_{N \ell }
        \nonumber \\
        &
        + [\gamma_N^{(Y)}]_{ae} C_{NN}^{e b cd} 
        + C_{NN}^{a e cd} [\gamma_N^{(Y)}]_{e b} 
        +[\gamma_N^{(Y)}]_{c e} C_{NN}^{a b ed} 
        + C_{NN}^{a b ce} [\gamma_N^{(Y)}]_{e d}\,,
        \\
    \nonumber\\
    \dot{C}^{ab\alpha\beta}_{Ne} &=  
        -2[Y^\dagger_\nu Y^{}_\nu]_{a b}C^{\alpha\beta}_{He}
        +2[Y_e^\dagger Y^{}_e]_{\alpha\beta} C^{ab}_{HN}
        +2[Y^\dagger_e Y^{}_\nu]_{\alpha b} C^{a\beta}_{HNe}
        +2[Y^\dagger_\nu Y^{}_e]_{a\beta}C^{*b\alpha}_{HNe}
        \nonumber \\
        &
        +[Y_\nu]_{\rho b}[Y_e]_{\sigma \beta}\Big(C^{*\sigma a \rho \alpha}_{\ell N \ell e}-C^{*\rho a \sigma \alpha}_{\ell N \ell e}\Big)
        +[Y^\dagger_\nu]_{a \rho }[Y^\dagger_e]_{\alpha \sigma}\Big(C^{\sigma b \rho \beta}_{\ell N \ell e}-C^{\rho b \sigma \beta}_{\ell N \ell e}\Big)
        \nonumber\\
        &
        -2[Y^\dagger_\nu]_{a \sigma} [Y^{}_\nu]_{\rho b}C^{\sigma \rho \alpha \beta}_{\ell e}
        -2[Y^\dagger_{e}]_{\alpha\rho}[Y^{}_{e}]_{\sigma \beta}C^{ab \rho\sigma }_{N \ell }
        \nonumber\\
        &
        + [\gamma_N^{(Y)}]_{ac} C_{Ne}^{c b \alpha\beta} 
        + C_{Ne}^{a c \alpha\beta} [\gamma_N^{(Y)}]_{c b} 
        +[\gamma_e^{(Y)}]_{\alpha \rho} C_{Ne}^{a b \rho\beta} 
        + C_{Ne}^{a b \alpha\rho} [\gamma_e^{(Y)}]_{\rho \beta}\,,
        \\
    \nonumber\\
    \dot{C}^{ab ij}_{Nu} &=  
        -2[Y^\dagger_u Y^{}_u]_{ij}C^{ab}_{HN}
        -2[Y^\dagger_\nu Y^{}_\nu]_{a b}C^{ij}_{Hu}
        -2[Y^\dagger_u]_{i k}[Y^{}_u]_{lj}C^{ab kl}_{Nq}
        -2[Y^\dagger_\nu]_{a \sigma} [Y^{}_\nu]_{\rho b}C^{\sigma \rho ij}_{\ell u}
        \nonumber \\
        &
        +[Y^{}_\nu]_{\rho b}[Y^\dagger_u]_{i k}C_{N\ell qu}^{a\rho k j}
        +[Y^\dagger_\nu]_{a \rho}[Y^{}_u]_{k j}C_{N\ell qu}^{*b \rho k i}
        \nonumber\\
        &
        + [\gamma_N^{(Y)}]_{ac} C_{Nu}^{c b ij} 
        + C_{Nu}^{a c ij} [\gamma_N^{(Y)}]_{c b} 
        +[\gamma_u^{(Y)}]_{i k} C_{Nu}^{a b kj} 
        + C_{Nu}^{a b ik} [\gamma_u^{(Y)}]_{k j} \,,
        \\
    \nonumber\\
    \dot{C}^{ab ij}_{Nd} &=  
        2[Y^\dagger_d Y^{}_d]_{ij} C^{ab}_{HN}
        -2[Y^\dagger_\nu Y^{}_\nu]_{a b}C^{ij}_{Hd}
        -2[Y^\dagger_d]_{ik}[Y^{}_d]_{lj}C^{ab kl}_{Nq}
        -2[Y^\dagger_\nu]_{a \sigma} [Y^{}_\nu]_{\rho b}C^{\sigma \rho ij}_{\ell d}
        \nonumber \\
        &
        {
        -[Y^\dagger_\nu]_{ a\sigma}[Y^\dagger_d]_{ik}\left(C^{\sigma b kj}_{\ell N q d}+C^{\sigma j k b}_{\ell d qN}\right)
        -[Y_\nu]_{\sigma b}[Y_d]_{kj}\left(C^{*\sigma a k i}_{\ell N q d}+C^{*\sigma i k a}_{\ell d qN}\right)
        }
        \nonumber\\
        &
        + [\gamma_N^{(Y)}]_{ac} C_{Nd}^{c b ij} 
        + C_{Nd}^{a c ij} [\gamma_N^{(Y)}]_{c b} 
        +[\gamma_d^{(Y)}]_{i k} C_{Nd}^{a b kj} 
        + C_{Nd}^{a b ik} [\gamma_d^{(Y)}]_{k j} \,,
        \\
    \nonumber\\
    \dot{C}^{a\alpha ij}_{Nedu} &=  
        {[Y_u]_{kj}[Y_e]_{\rho\alpha}(C^{*\rho a ki}_{\ell Nqd}
        +C^{*\rho i k a}_{\ell d qN})}
        +[Y^{}_e]_{\rho \alpha}[Y^\dagger_d]_{ik}C^{a\rho k j}_{N\ell q u}
        +[Y^\dagger_\nu]_{a\sigma}[^{}Y_u]_{kj}C^{\sigma \alpha ik}_{\ell e dq}
        \nonumber \\
        &
        {-}[Y^\dagger_\nu]_{a\sigma}[Y^\dagger_d]_{ik}\Big(C^{(1)\sigma \alpha kj}_{\ell e qu} -12C^{(3)\sigma \alpha kj}_{\ell e qu}\Big)
        +2[Y^\dagger_d Y^{}_u]_{ij}C^{a \alpha}_{HNe}
        +2[Y^\dagger_\nu Y^{}_e]_{a\alpha} C^{*ji}_{Hud}
        \nonumber\\
        &
        + [\gamma_N^{(Y)}]_{ac} C_{Nedu}^{c \alpha ij} 
        + C_{Nedu}^{a \beta ij} [\gamma_e^{(Y)}]_{\beta \alpha} 
        +[\gamma_d^{(Y)}]_{i k} C_{Nedu}^{a \alpha kj} 
        + C_{Nedu}^{a \alpha ik} [\gamma_u^{(Y)}]_{k j} \,.
\end{align}

\subsubsection[$(\bar{L}L)(\bar{R}R)$]{$\boldsymbol{(\bar{L}L)(\bar{R}R)}$}
\begin{align}
    \dot{C}^{a b\alpha\beta }_{N\ell} &=  
        [Y^\dagger_\nu]_{a\beta}[\xi_N]_{\alpha b}
        + [Y_\nu]_{\alpha b}[\xi^*_N]_{\beta a}
        +[(Y^{}_\nu Y^\dagger_\nu)-(Y_e Y^\dagger_e)]_{\alpha\beta} C^{ab}_{HN}
        -2[Y^\dagger_\nu Y^{}_\nu]_{a b}C^{(1)\alpha\beta}_{H\ell}
        \nonumber \\
        &
        +[Y^{}_\nu]_{\rho b}[Y^\dagger_\nu]_{c\beta} C^{a c \alpha \rho}_{N \ell}
        +[Y^{}_\nu]_{\alpha c}[Y^\dagger_\nu]_{a \rho} C^{c b \rho \beta}_{N \ell}
        - 2[Y^{}_{\nu}]_{\sigma b}[Y^\dagger_{\nu}]_{a\rho} C^{\alpha \sigma \rho \beta}_{\ell \ell}
        -4[Y^{}_{\nu}]_{\rho b}[Y^\dagger_{\nu}]_{a\sigma} C^{\alpha \beta \sigma\rho }_{\ell \ell}
        \nonumber\\
        &
        -[Y^{}_e]_{\alpha \rho}[Y_e^\dagger]_{\sigma \beta} C^{ ab \rho\sigma}_{Ne}
        -2[Y^\dagger_\nu]_{\beta c}[Y^{}_\nu]_{\alpha d}(C^{abdc}_{NN}+C^{acdb}_{NN})
        \nonumber\\
        &
        +[Y^\dagger_\nu]_{a \rho}[Y_e^\dagger]_{\sigma \beta}\left(C^{\alpha b \rho \sigma}_{\ell N\ell e}
        +\frac12C^{\rho b \alpha \sigma}_{\ell N\ell e}\right)
        +[Y_\nu]_{\rho b}[Y_e]_{\alpha \sigma}\left(C^{*\beta a \rho \sigma}_{\ell N\ell e}
        +\frac12C^{*\rho a \beta \sigma}_{\ell N\ell e}\right)
        \nonumber\\
        &
        + [\gamma_N^{(Y)}]_{ac} C_{N\ell}^{c b \alpha\beta} 
        + C_{N\ell}^{a c \alpha\beta} [\gamma_N^{(Y)}]_{c b} 
        +[\gamma_\ell^{(Y)}]_{\alpha \rho} C_{N\ell}^{a b \rho\beta} 
        + C_{N\ell}^{a b \alpha\rho} [\gamma_\ell^{(Y)}]_{\rho \beta}\,,
        \\
    \nonumber\\
    \dot{C}^{abij}_{Nq} &=  
        [(Y^{}_uY^\dagger_u)-(Y_d Y^\dagger_d)]_{ij} C^{ab}_{HN}
        -2[Y^\dagger_\nu Y^{}_\nu]_{a b}C^{(1)ij}_{Hq}
        -2[Y^\dagger_\nu]_{a \sigma} [Y^{}_\nu]_{\rho b}C^{(1)\,\sigma \rho ij}_{\ell q}
        \nonumber \\
        &
        -[Y^{}_u]_{ik}[Y^\dagger_u]_{lj}C^{ab kl}_{Nu}
        -[Y^{}_d]_{ik}[Y^\dagger_d]_{lj}C^{ab kl}_{Nd}
        -\frac{1}{2}[Y^{}_\nu]_{\rho b}[Y^\dagger_u]_{kj}C^{a \rho i k}_{N\ell qu}
        -\frac{1}{2}[Y^\dagger_\nu]_{a \rho }[Y^{}_u]_{ik}C^{*b\rho j k}_{N\ell qu}
        \nonumber\\
        &
        +
        {
        \frac{1}{2}[Y^\dagger_\nu]_{a\rho}[Y^\dagger_d]_{kj}(C^{\rho b i k}_{\ell N qd}-2C^{\rho k i b}_{\ell d qN})
        +\frac{1}{2}[Y_\nu]_{\rho b}[Y_d]_{ik}(C^{*\rho a jk}_{\ell N qd}-2C^{*\rho k ja}_{\ell d qN})
        }
        \nonumber\\
        &
        + [\gamma_N^{(Y)}]_{ac} C_{Nq}^{c b ij} 
        + C_{Nq}^{a c ij} [\gamma_N^{(Y)}]_{c b} 
        +[\gamma_q^{(Y)}]_{i k} C_{Nq}^{a b kj} 
        + C_{Nq}^{a b ik} [\gamma_q^{(Y)}]_{k j} \,,
        \\
    \nonumber\\
    \dot{C}^{\alpha\beta ij}_{\ell u} &=  
        [Y^{}_\nu Y^\dagger_\nu]_{\alpha\beta}C^{ij}_{Hu}
        -[Y^{}_\nu]_{\alpha a} [Y^\dagger_{\nu}]_{b \beta}C^{abij}_{Nu} 
        -\frac{1}{2} [Y^{}_\nu]_{\alpha a}[Y_u^{\dagger}]_{ik} C^{a\beta kj}_{N\ell qu}
        -\frac{1}{2} [Y^\dagger_\nu]_{a\beta}[Y^{}_u]_{kj} C^{*a \alpha ki}_{N\ell qu}\,,
        \\
    \nonumber\\
    \dot{C}^{\alpha\beta ij}_{\ell d} &=  
        [Y^{}_\nu Y^\dagger_\nu]_{\alpha\beta}C^{ij}_{Hd}
        -[Y^{}_\nu]_{\alpha a} [Y^\dagger_{\nu}]_{b \beta } C^{abij}_{Nd} 
        +\frac{1}{2} [Y^{\dagger}_\nu]_{a \beta } [Y_d^{\dagger}]_{ik}(C^{\alpha a kj}_{\ell N qd} -2C^{\alpha j k a}_{\ell d qN}) 
        \nonumber \\
        &
        +
        {
        \frac{1}{2} [Y_\nu]_{\alpha c} [Y_d]_{kj}(C^{*\beta c ki}_{\ell N qd} -2C^{*\beta i kc}_{\ell d qN})\,,
        }
        \\
    \nonumber\\    
    \dot{C}^{\alpha\beta \delta\gamma}_{\ell e} &=  
        [Y^{}_\nu Y^\dagger_\nu]_{\alpha\beta}C^{\delta \gamma}_{He}
        -[Y^{}_\nu]_{\alpha a} [Y^\dagger_{\nu}]_{ b\beta}C^{ab\delta\gamma}_{Ne} 
        \nonumber \\
        &
        +\frac{1}{2} [Y_\nu^{\dagger}]_{a\beta} [Y_e^{\dagger}]_{\delta \rho} \Big(C^{\alpha a \rho \gamma}_{\ell N \ell e} +2C^{\rho a \alpha\gamma}_{\ell N \ell e}\Big)
        +\frac{1}{2} [Y_\nu]_{\alpha a} [Y_e]_{\rho\gamma}\Big(C^{* \beta a \rho \delta}_{\ell N \ell e}+2C^{* \rho a \beta \delta}_{\ell N \ell e}\Big) \,.
\end{align}

\subsubsection[$(\bar{L}L)(\bar{L}L)$]{$\boldsymbol{(\bar{L}L)(\bar{L}L)}$}

\begin{align}
    \dot{C}^{\alpha\beta\gamma\delta}_{\ell \ell} &=  
         \frac{1}{2}[Y^{}_\nu Y^\dagger_\nu]_{\alpha \beta} \big(C^{(1)\gamma \delta}_{H\ell}+C^{(3)\gamma \delta}_{H\ell}\big)
         +\frac{1}{2}[Y^{}_\nu Y^\dagger_\nu]_{\gamma \delta} \big(C^{(1)\alpha \beta}_{H\ell} +C^{(3)\alpha \beta}_{H\ell} \big)
        \nonumber \\
        &
         -[Y^{}_\nu Y^\dagger_\nu]_{\alpha \delta}C^{(3)\gamma \beta}_{H\ell}
         -[Y^{}_\nu Y^\dagger_\nu]_{\gamma \beta }C^{(3)\alpha \delta}_{H\ell}
        -\frac{1}{2}[Y^{}_\nu]_{\alpha a}[Y^\dagger_{\nu}]_{b \beta } C^{ab \gamma \delta}_{N\ell} 
        -\frac{1}{2}[Y^{}_\nu]_{\gamma a} [Y^\dagger_{\nu}]_{b \delta} C^{ab\alpha\beta}_{N\ell}
        \nonumber\\ 
        &
        -\frac{1}{4} [Y_\nu^{\dagger}]_{a \beta } [Y_e^{\dagger}]_{\rho \delta} C^{\alpha a \gamma \rho}_{\ell N \ell e}  
        -\frac{1}{4} [Y_\nu^{\dagger}]_{a \delta } [Y_e^{\dagger}]_{\rho \beta} C^{\gamma a \alpha \rho}_{\ell N \ell e} 
        +\frac{1}{4} [Y_\nu^{\dagger}]_{ a \delta } [Y_e^{\dagger}]_{\rho\beta} C^{\alpha a \gamma \rho}_{\ell N \ell e} 
        \nonumber \\
         &
        +\frac{1}{4} [Y_\nu^{\dagger}]_{a \beta } [Y_e^{\dagger}]_{\rho \delta} C^{\gamma a \alpha \rho}_{\ell N \ell e} 
        - \frac{1}{4} [Y_\nu]_{\alpha a}[Y_e]_{\gamma \rho} C^{*\beta a \delta \rho}_{\ell N \ell e}
        - \frac{1}{4} [Y_\nu]_{\gamma a}[Y_e]_{\alpha \rho} C^{*\delta a \beta \rho}_{\ell N \ell e}
         \nonumber\\
         &
         + \frac{1}{4} [Y_\nu]_{\gamma a }[Y_e]_{\alpha \rho} C^{*\beta a \delta \rho}_{\ell N \ell e}
         +\frac{1}{4} [Y_\nu]_{\alpha a }[Y_e]_{\gamma \rho} C^{*\delta a \beta \rho}_{\ell N \ell e}\,,
        \\
    \nonumber\\
    \dot{C}^{(1)\alpha\beta ij}_{\ell q} &=  
        [Y^{}_\nu Y^\dagger_\nu]_{\alpha\beta}C^{(1)ij}_{Hq}
        -[Y^{}_\nu]_{\alpha a} [Y^\dagger_{\nu}]_{b \beta }C^{abij}_{Nq}
        +\frac{1}{4} [Y^{}_\nu]_{\alpha a}[Y_u^{\dagger}]_{kj}C^{a\beta ik}_{N\ell qu}
        +\frac{1}{4} [Y^\dagger_\nu]_{a\beta}[Y^{}_u]_{ik}C^{*a \alpha jk}_{N\ell qu}
        \nonumber \\
        &
        -\frac{1}{4} [Y^{\dagger}_\nu]_{a \beta } [Y_d^{\dagger}]_{kj}\Big(C^{\alpha a i k}_{\ell N qd} +C^{\alpha k i a}_{\ell d qN}\Big)
        -\frac{1}{4} [Y_\nu]_{\alpha a } [Y_d]_{ik}\Big(C^{*\beta a j k}_{\ell N qd}+C^{*\beta k j a}_{\ell d qN}\Big)\,,
        \\
    \nonumber\\
    \dot{C}^{(3)\alpha\beta ij}_{\ell q} &=  
        -[Y^{}_\nu Y^\dagger_\nu]_{\alpha\beta}C^{(3)ij}_{Hq}
        +\frac{1}{4} [Y^{}_\nu]_{\alpha a}[Y_u^{\dagger}]_{kj}C^{a\beta ik}_{N\ell qu}
        +\frac{1}{4} [Y^\dagger_\nu]_{a\beta}[Y^{}_u]_{ik}C^{*a \alpha jk}_{N\ell qu}
        \nonumber \\
        &
        +\frac{1}{4} [Y^{\dagger}_\nu]_{a \beta } [Y_d^{\dagger}]_{kj}\Big(C^{\alpha a i k}_{\ell N qd} +C^{\alpha k i a}_{\ell d qN}\Big)
        +\frac{1}{4} [Y_\nu]_{\alpha a } [Y_d]_{ik}\Big(C^{*\beta a j k}_{\ell N qd}+C^{*\beta k j a}_{\ell d qN}\Big)\,.
\end{align}

\subsubsection[$(\bar{L}R)(\bar{L}R)$]{$\boldsymbol{(\bar{L}R)(\bar{L}R)}$}

\begin{align}
    \dot{C}^{\alpha a \beta \delta}_{\ell N \ell e} &=  
        -2[Y_\nu]_{\alpha a}[\xi_e]_{\beta \delta}
        -2[Y_e]_{\beta \delta} [\xi_N]_{\alpha a}
        +4 [Y_\nu]_{\beta c}[Y_e]_{\sigma \delta}C^{ca \alpha \sigma}_{N\ell}
        +4[Y_e]_{\alpha \sigma} [Y_\nu]_{\rho a}C^{\beta \rho \sigma \delta}_{\ell e}
        \nonumber \\
        &
        -4\Big( [Y_e]_{\sigma \delta} [Y_\nu]_{\rho a} {
        -[Y_e]_{\rho \delta} [Y_\nu]_{\sigma a}}\Big) C^{\alpha \rho \beta \sigma}_{\ell \ell}
        +4 \Big([Y_e]_{\sigma \delta} [Y_\nu]_{\rho a} {
        -[Y_e]_{\rho \delta} [Y_\nu]_{\sigma a}}\Big) C^{\beta \rho \alpha \sigma}_{\ell \ell}       
        \nonumber\\
        &
        -4\Big([Y_\nu]_{\alpha c}[Y_e]_{\beta \sigma}-[Y_\nu]_{\beta c}[Y_e]_{\alpha \sigma} \Big) { C^{c \delta \sigma a}_{Ne}}
        \nonumber\\
        &
        + [\gamma_\ell^{(Y)}]_{\alpha \rho} C_{\ell N \ell e}^{\rho a\beta \delta} 
        + C_{\ell N\ell e}^{\alpha b \beta \delta } [\gamma_N^{(Y)}]_{b a} 
        +[\gamma_\ell^{(Y)}]_{\beta \rho } C_{\ell N \ell e}^{\alpha a \rho \delta } 
        + C_{\ell N\ell e}^{\alpha a \beta \rho} [\gamma_e^{(Y)}]_{\rho \delta}
        \nonumber\\
        &
        {
        -2g' [Y_\nu]_{\alpha a} C_{eB}^{\beta\delta}
        -6g  [Y_\nu]_{\alpha a} C_{eW}^{\beta\delta}
        +4g' [Y_\nu]_{\beta a} C_{eB}^{\alpha \delta}
        +12g [Y_\nu]_{\beta a} C_{eW}^{\alpha \delta}
        }
        \nonumber\\
        &
        {
        -6g' [Y_e]_{\alpha a} C_{NB}^{\beta\delta}
        -6g [Y_e]_{\alpha a} C_{NW}^{\beta\delta}
        +12g' [Y_e]_{\beta a} C_{NB}^{\alpha \delta}
        +12g [Y_e]_{\beta a} C_{NW}^{\alpha \delta}\,,
        }
        \\
    \nonumber\\
    \dot{C}^{\alpha a ij}_{\ell Nqd} &=  
        -2[Y_d]_{ij} [\xi_N]_{\alpha a}
        -2[Y_\nu]_{\alpha a}[\xi_d]_{ij}
        \nonumber\\
        &
        -4[Y_\nu]_{\alpha b}[Y_d]_{ik} C^{bakj}_{Nd}
        +4[Y_e]_{\alpha \rho}[Y_u]_{ik}C^{*a\rho jk}_{Nedu}
        -4 [Y_d]_{kj}[Y_\nu]_{\sigma a}C^{(1)\alpha \sigma ij}_{\ell q}
        \nonumber\\
        &
        +12 [Y_d]_{kj}[Y_\nu]_{\sigma a}C^{(3)\alpha \sigma ij}_{\ell q}
        -2[Y_d]_{kj}[Y_u]_{il}C^{*a\alpha kl}_{N\ell qu}
        -2 [Y^{}_\nu]_{\rho a}[Y_e]_{\alpha \sigma }C^{*\rho \sigma ji}_{\ell e dq}
        \nonumber\\
        &
        + [\gamma_\ell^{(Y)}]_{\alpha \rho} C_{\ell Nqd}^{\rho ai j} 
        + C_{\ell N q d}^{\alpha bij } [\gamma_N^{(Y)}]_{b a} 
        +[\gamma_q^{(Y)}]_{ik } C_{\ell N q d}^{\alpha a k j } 
        + C_{\ell N q d}^{\alpha a i k} [\gamma_d^{(Y)}]_{k j} 
        \nonumber\\
        &
        -\frac{2}{3} g' [Y_d]_{ij}C^{\alpha a}_{NB} -6 g [Y_d]_{ij}C^{\alpha a}_{NW}- 2 g' [Y_\nu]_{\alpha a}C^{ij}_{dB} -6 g [Y_\nu]_{\alpha a}C^{ij}_{dW}\,,
        \\
    \nonumber\\
    \dot{C}^{\alpha ji a}_{\ell dqN} &=  
        -4[Y_\nu]_{\alpha b}[Y_d]_{ik} C^{bakj}_{Nd}
        +4[Y_e]_{\alpha \rho}[Y_u]_{ik}C^{*a\rho jk}_{Nedu}
        -4 [Y_{\nu}]_{\alpha b}[Y_d]_{kj}C^{baik}_{Nq}
        \nonumber\\
        &
        -4 [Y_d]_{kj}[Y_\nu]_{\sigma a}C^{(1)\alpha \sigma ij}_{\ell q}
        +12 [Y_d]_{kj}[Y_\nu]_{\sigma a}C^{(3)\alpha \sigma ij}_{\ell q}
        -4[Y_d]_{ik}[Y_\nu]_{\sigma a}C^{\alpha \sigma kj}_{\ell d}
        \nonumber\\
        &
        + [\gamma_\ell^{(Y)}]_{\alpha \rho} C_{\ell dqN}^{\rho ji a} 
        + C_{\ell d q N}^{\alpha kia } [\gamma_d^{(Y)}]_{k j} 
        +[\gamma_q^{(Y)}]_{ik } C_{\ell dqN}^{\alpha j k a } 
        + C_{\ell d q N}^{\alpha j i b} [\gamma_N^{(Y)}]_{b a} 
        \nonumber\\
        &
        -\frac{4}{3} g' [Y_d]_{ij}C^{\alpha a}_{NB} -12 g [Y_d]_{ij}C^{\alpha a}_{NW}- 4 g' [Y_\nu]_{\alpha a}C^{ij}_{dB} -12 g [Y_\nu]_{\alpha a}C^{ij}_{dW}
        \,,
        \\
    \nonumber\\
    \dot{C}^{(1)\alpha\beta ij}_{\ell e qu} &=  
        -2  [Y_\nu]_{\alpha a} [Y_d]_{ik} C^{a\beta kj}_{Nedu}
        {
        +2 [Y_\nu]_{\alpha a} [Y_e]_{\sigma\beta} C_{N\ell qu}^{a\sigma ij}
        }\,,
        \\
    \nonumber\\
    \dot{C}^{(3)\alpha\beta ij}_{\ell e qu} &=  
        \frac{1}{2} [Y_\nu]_{\alpha a} [Y_d]_{ik} C^{a\beta kj}_{Nedu}.
\end{align}

\subsubsection[$(\bar{L}R)(\bar{R}L)$]{$\boldsymbol{(\bar{L}R)(\bar{R}L)}$}

\begin{align}
    \dot{C}^{a \alpha ij}_{N\ell q u} &=  
        -2[Y_u]_{ij} [\xi^*_N]_{ \alpha a}
        -2[Y^\dagger _\nu]_{ a\alpha}[\xi_u]_{ij}
        +2 [Y^\dagger_\nu]_{c \alpha }[Y_u]_{ik} C^{ac k j}_{Nu}
        +2[Y^{}_d]_{ik}[Y^\dagger_e]_{\sigma\alpha}C^{a \sigma kj}_{Ne du} 
        \nonumber \\
        &
        -2[Y^{}_u]_{kj}[Y^\dagger_\nu]_{b \alpha }C^{ab i k}_{Nq}
        +2[Y^{}_u]_{ik}[Y^\dagger_\nu]_{a\sigma}C^{\sigma \alpha kj}_{\ell u}
        + 2[Y^{}_u]_{kj}[Y^\dagger_\nu]_{a\sigma}C^{(1)\sigma \alpha ik}_{\ell q}
        \nonumber \\
        &
        + 6[Y^{}_u]_{kj}[Y^\dagger_\nu]_{a\sigma}C^{(3)\sigma \alpha ik}_{\ell q}
        +2[Y^\dagger_\nu]_{a\sigma}[Y^\dagger_e]_{\rho \alpha}C^{(1)\sigma \rho ij}_{\ell e qu}
        -2[Y_d]_{ik}[Y_u]_{lj}\Big(C^{*\alpha a lk}_{\ell N qd}-\frac12 C^{*\alpha k lN}_{\ell d q N}\Big)
        \nonumber \\
        &
        + [\gamma_N^{(Y)}]_{a b} C_{N \ell q u}^{b \alpha i j} 
        + C_{N \ell qu}^{a \beta ij } [\gamma_\ell ^{(Y)}]_{\beta \alpha } 
        +[\gamma_q^{(Y)}]_{ik } C_{N \ell q u}^{a \alpha k j } 
        + C_{N \ell q u}^{a \alpha i k } [\gamma_u^{(Y)}]_{k j}\,,
        \\
    \nonumber\\
    \dot{C}^{\alpha\beta ij}_{\ell e dq} &=  
        2 [Y^{}_\nu]_{\alpha a} [Y^\dagger_u]_{kj} C^{a\beta ik}_{Nedu}
        {
        +[Y_{\nu}]_{\alpha a}[Y_e]_{\rho \beta}C^{*\rho i ja}_{\ell d q N}
        }
        \,.
\end{align}


\bibliographystyle{JHEP} 
\bibliography{biblio}

\end{document}